\documentclass[12pt,epsfig]{article}

\usepackage{epsfig}

\makeatletter
\renewcommand{\arraystretch}{1.5}

\newlength{\abstwidth}
\def\varep{\varepsilon}

\def\ep{e^{i\phi}}
\def\emp{e^{-i\phi}}
\def\etp{e^{2i\phi}}
\def\emtp{e^{-2i\phi}}

\def\repp{\rho_1^{++}}
\def\rtpp{\rho_2^{++}}
\def\renn{\rho_1^{00}}
\def\rtnn{\rho_2^{00}}
\def\repn{|\rho_1^{+0}|}
\def\rtpn{|\rho_2^{+0}|}  
\def\repm{|\rho_1^{+-}|}
\def\rtpm{|\rho_2^{+-}|}

\def\eip{e^{i\tilde{\phi}}}
\def\emip{e^{-i\tilde{\phi}}}

\def\emtip{e^{-2i\tilde{\phi}}}

\def\eipe{e^{i\tilde{\phi}_1}}

\def\etipe{e^{2i\tilde{\phi}_1}}

\def\efipe{e^{4i\tilde{\phi}_1}}

\def\e3ipe{e^{3i\tilde{\phi}_1}}
\def\em3ipe{e^{-3i\tilde{\phi}_1}}

\def\stt{\sigma_{TT}}

\def\tts{\tau_{TS}}

\def\fp{e^{i\phi}}
\def\fm{e^{-i\phi}}
\def\tfp{e^{2i\phi}}
\def\tfm{e^{-2i\phi}}

\begin{document}
\renewcommand{\arraystretch}{1.3}

\begin{center}

{\LARGE \bf Two-photon mediated resonance production in $e^+e^-$ collisions:
  cross sections and density matrices}

\vspace{1.0cm}

{\Large F.\ A.\ Berends$^a$ and R.\ van Gulik$^{a,b}$} \\[2ex]

{\it   
${}^a$Instituut-Lorentz, 
University of Leiden, P.O. Box 9506, 2300 RA Leiden,  
The Netherlands\\[2ex]
${}^b$NIKHEF, P.O. Box 41882, 1009 DB Amsterdam, The Netherlands
}

\vspace{1.0cm}

\end{center}

 
\begin{center}
{\bf Abstract}\\[2ex]
\begin{minipage}{\abstwidth}
Earlier described model amplitudes are used in this paper to evaluate both
cross sections and density matrices for two-photon mediated resonance
production in $e^+e^-$ collisions. All 25 $q\bar{q}$ low-lying $^1S_0$,
$^3P_J$ and $^1D_2$ resonances can thus be treated. Two independent methods 
are described to obtain the resonance production density matrices and
cross sections. These density matrices combined with a resonance 
decay density matrix give the detailed angular distributions of the resonance
decay products. For two particular decays, 
$\chi_{c2},\chi_{c1}\to\gamma J/\psi$ the 
details are given. Several numerical results are presented as well.
\end{minipage}
\end{center}


\vspace{1.0cm}

\section{Introduction}

When performing electron-positron collision experiments, there will always be
data originating from (virtual) two-photon collisions. In the latter
process single resonance production occurs. When present and
future experiments collect more and more
statistics it becomes worthwile to have theoretical predictions available for 
observables, which gradually become experimentally accessible. 

In a previous paper \cite{SchuBerGul}, 
hereafter called I, a specific class of such observables
namely form factors were studied. In this paper other observables will be
focussed upon, such as azimuthal dependences in cross sections and resonance
production density matrices. The latter quantity combined with a resonance
decay density matrix will give the distribution of the decay products of the 
resonance.

In order to study the above observables one needs a model for resonance
production. In I such a model was described starting from heavy quarks in a 
non-relativistic bound state. The predicted two-photon decay widths of 
resonances $R$ turn out to be in reasonable agreement with experiment.
Once certain modifications are made, also the widths for light quark mesons 
become reasonable. Therefore the model of I is chosen as basis to predict
observables for the 25 low-lying $^1S_0$, $^3P_J$ and $^1D_2$ resonances.
In paper I analytic amplitudes for $\gamma\gamma \to R$ are
presented and expressions for form factors are derived. A number of 
numerical results for cross sections and $Q_i^2$ dependences can also be found
there. 

In the present paper numerical studies of azimuthal distributions will be
given. Their characteristic features can be understood from various 
analytic expressions. Whereas this discussion is a simple extension of the
results of I, the evaluation of production density matrices requires
substantial changes in calculational techniques. Moreover, rotations are
required between reference frames favoured by theory and experiment.
A discussion of two different evaluation methods will be given in this paper.
Finally, in order to make a link to the angular distribution of resonance
decay products, also one specific decay mode for two particular resonances
will be 
discussed.

Summarizing, the purpose of the present paper is to discuss theoretical 
evaluation methods for resonance production density matrices and to present
results thereof. Thus the full predictions of the resonance production model
of I are in this way completed.

The paper is organized as follows. Section 2 gives the
relevant amplitudes for resonance production. The direct method of the
evaluation of the observables is discussed in section 3, whereas section 4 
gives an extension of the BGMS \cite{BGMS} formalism required for the 
evaluation of density matrices. After discussing a specific decay model 
for 
two
resonances in section 5, numerical results for several 
quantities are presented in section 6.


\section{The starting point}

In this section the basic ingredients of the evaluation are listed, such
as kinematics and matrix elements following from 
the model of I.

The two-photon mediated production of a resonance $R$ in 
electron-positron collisions is the reaction
\renewcommand{\arraystretch}{1.8}
\begin{equation}
  \begin{array}{l}
    e^+(p_1)+e^-(p_2)\to e^+(p_1')+e^-(p_2')+\gamma(k_1)+\gamma(k_2) 
    \\
    \to
    e^+(p_1')+e^-(p_2')+R(p_R).
  \end{array}
  \label{ReacForm}
\end{equation}
The four-momenta
$p_1=(E_1,\vec{p}_1)$ and $p_2=(E_2,\vec{p}_2)$ correspond to the 
incoming positron and
electron respectively\footnote{
We use the standard metric $\mbox{diag}(g^{\mu\nu})
=(+1,-1,-1,-1)$. The totally anti-symmetric Levi-Civita tensor is
defined by $\varepsilon_{0123}=+1$. 
}, whereas
$p_1'=(E_1',\vec{p'}_1)$ and $p_2'=(E_2',\vec{p'}_2)$ are those 
of the outgoing positron and electron.
The four-momenta of the intermediate photons $k_1=(\omega_1,\vec{k}_1)$ and
$k_2=(\omega_2,\vec{k}_2)$ are related to the external four-momenta by
\begin{equation} 
  k_i =  p_i -p_i'.
\end{equation}
These two photons subsequently react to form a resonance with 
four-momentum $p_R$
\begin{equation}
  p_R=k_1+k_2=p_1+p_2-p_1'-p_2'.
\end{equation}
The virtuality $Q_i$ of an intermediate photon is defined 
by
\begin{equation}
  Q_i^2 \equiv -k_i^2 = -(p_i-p_i')^2.
  \label{virtdef}
\end{equation}
As the virtual photons are
space-like, the sign in (\ref{virtdef}) 
makes
the 
virtualities $Q_i$  
real.  

The invariant mass of the two-photon system is given by
\begin{equation}
  W
  =p_R^2=(k_1+k_2)^2.
  \label{wggdef}
\end{equation}

The total available energy $\sqrt{s}$
follows from
\begin{equation}
  s = (p_1 + p_2)^2.
  \label{ForSDef}
\end{equation}
For a typical two photon event $W$ is only a small
fraction of $\sqrt{s}$.

Some additional invariants can be defined
\begin{eqnarray}
  s' & = & (p_1'+p_2')^2, \\
  u & = & (p_1-p_2')^2 ,\\
  u' & = & (p_2 - p_1')^2 \label{ForUpDef},
\end{eqnarray} 
and as equivalents of $Q_i^2$
\begin{eqnarray}
  t & = & (p_1-p_1')^2, \\
  t'& = & (p_2-p_2')^2. \label{ForTpDef}
\end{eqnarray}
These invariants satisfy 
\begin{equation}
  s+s'+t+t'+u+u'=W^2+8m_e^2,
\end{equation}
where $m_e$ is the electron mass.

In the lab-frame 
we chose the following parametrization
\renewcommand{\arraystretch}{1.2}
\begin{equation}
   \label{labpara}
   \begin{array}{lcl}
      p_1 & = & (E_b,0,0,-P_b), \\
      p_2 & = & (E_b,0,0,P_b), \\
      p_1' & = & (E_1',|\vec{p'}_1|\sin \theta_1 \cos \phi_1,|\vec{p'}_1|\sin \theta_1 \sin \phi_1,
                            -|\vec{p'}_1| \cos \theta_1), \\
      p_2' & = & (E_2',|\vec{p'}_2|\sin \theta_2 \cos \phi_2,|\vec{p'}_2|\sin \theta_2 \sin \phi_2,
                            |\vec{p'}_2| \cos \theta_2), \\
      p_R & = & (E_R,|\vec{p}_R|\sin \theta_R \cos \phi_R,|\vec{p}_R|\sin \theta_R \sin \phi_R,
                            |\vec{p}_R| \cos \theta_R). 
   \end{array}
\end{equation}
This means that the $z$ axis is along the direction of the incoming electron.
The $x$ and $y$ axes are chosen such that the $xz$-plane is the accelerator
ring plane and the $y$ axis is perpendicular to that plane and is pointing
upwards.
In these formulae $E_b$ is the energy of the incoming leptons and 
$P_b$ their momentum.
The choice of parameters in (\ref{labpara}) is such that 
we have defined the $z$-components of the positron with an explicit minus
sign.

This parametrization 
leads to 
explicit 
expressions for the invariants
\begin{equation}
  s = 4 E_b^2,
\end{equation}
\begin{equation}
  Q_i^2 = 2\left(E_bE_i'-P_b|\vec{p'}_i|\cos \theta_i-m_e^2\right) 
  \approx 2E_bE_i'(1-\cos\theta_i)=4E_bE_i'\sin^2\left(\frac{\theta_i}{2}
  \right),
  \label{Qi2def}
\end{equation}
\renewcommand{\arraystretch}{2.0}
\begin{equation}
  \begin{array}{lcl}
  W & = & \sqrt{2m_e^2+4(E_b-E_1')(E_b-E_2')-2E_1'E_2'
  -2|\vec{p'}_1||\vec{p'}_2|\cos \theta_{12}}  \\
  & \approx &
  2\sqrt{\omega_1\omega_2-E_1'E_2'\cos^2 \frac{\theta_{12}}{2}}.
  \label{WggDef}
  \end{array}
\end{equation}
The approximations are valid in the limit $E_b>E_i'>>m_e$.
In formula (\ref{WggDef}) the angle $\theta_{12}$ 
is the angle between the outgoing leptons in the lab-frame
\renewcommand{\arraystretch}{3.0}
\begin{equation}
  \cos \theta_{12} = -\cos \theta_1 \cos\theta_2 - \sin\theta_1\sin\theta_2
  \cos(\phi_1-\phi_2+\pi).
  \label{theta12def}
\end{equation}

The propagators of the intermediate photons will 
appear as a product of the photon 
virtualities in the denominator of
the matrix elements.
This is the most dominant dependence on the photon virtualities.
From (\ref{Qi2def}) it follows that the
photons are most likely radiated 
with low energy under a small angle with respect to
the incoming lepton.

When the incoming lepton scatters over a large 
enough angle, so that it is detected in one
of the (forward) detectors, this lepton 
is referred to as a tagged lepton. From the detected lepton
the four-momentum, and thus the virtuality, of the associated 
intermediate photon can
be reconstructed.

The two methods of calculation both start from the cross section formula
\begin{equation}
  \mbox{d}\sigma = \frac{(2\pi)^4
  \delta^{(4)}(k_1+k_2-P)}
  {4\sqrt{(p_1\cdot p_2)^2-m_e^4}} \frac{1}{4} \sum |{\cal M}|^2
  \frac{\mbox{d}^3\vec{p'}_1}{(2\pi)^32E_1'}
  \frac{\mbox{d}^3\vec{p'}_2}{(2\pi)^32E_2'}
  \mbox{d} \Gamma,
  \label{SigmaDef}
\end{equation}
with $ \mbox{d} \Gamma$ the invariant phase space element for the 
resonance.
The amplitude ${\cal M}$ has the structure
\begin{equation}
  {\cal M}=\frac{e^2}{tt'}j_1^\mu j_2^\nu M_{\mu\nu},
  \label{GeneralMatEl}
\end{equation}
with
\begin{eqnarray}
  j_1^\mu & = & j_1^\mu(\lambda_1,\lambda_1') =  \bar{v}_{\lambda_1}
  (p_1)\gamma^\mu v_{\lambda_1'}(p_1'), 
  \label{ForLepCur1}
  \\
  j_2^\mu & = & j_2^\mu(\lambda_2,\lambda_2') =  \bar{u}_{\lambda_2'}
  (p_2')\gamma^\mu u_{\lambda_2}(p_2),
  \label{ForLepCur2}
\end{eqnarray}
or explicitly in the model described (cf.\ also \cite{Papers}) in 
I\footnote{In these equations we have introduced
the shorthand notation $\varepsilon(p,q,r,s)\equiv 
\varepsilon^{\alpha\beta\gamma\delta}p_\alpha q_\beta r_\gamma s_\delta$.}
\renewcommand{\arraystretch}{1.4}
\begin{equation}
    {\cal M}(^1S_0;\lambda_1,\lambda_2,\lambda_1',\lambda_2')  =  
    \frac{c_1e^2}{tt'} \varepsilon \left[ k_1,k_2,j_1(\lambda_1,\lambda_1'),
    j_2(\lambda_2,\lambda_2')\right],
    \label{For1S0TotMatEl}
\end{equation}
\begin{equation}
  \begin{array}{lcl}
    {\cal M}(^3P_0;\lambda_1,\lambda_2,\lambda_1',\lambda_2') & = &     
    \frac{c_2e^2}{tt'} \left( \left[ j_1(\lambda_1,\lambda_1') \cdot 
    j_2(\lambda_2,\lambda_2') 
    k_1 \cdot k_2  \right. \right. \\
    && \left. - j_1(\lambda_1,\lambda_1') 
    \cdot k_2 j_2(\lambda_2,\lambda_2')
    \cdot k_1 \right] (W^2+ k_1 \cdot k_2) \\
    && \left.  - 
    j_1(\lambda_1,\lambda_1')\cdot j_2(\lambda_2,\lambda_2') t t' \right), 
  \end{array}
\end{equation}
\begin{equation}
  \begin{array}{lcl}
    {\cal M}(^3P_1;\lambda_1,\lambda_2,\lambda_1',\lambda_2',\lambda_R) & = & 
    \frac{c_3e^2}{tt'} \left( t\varepsilon \left[ \varepsilon^*(\lambda_R),
    j_1(\lambda_1,\lambda_1'), j_2(\lambda_2,\lambda_2') ,k_2 
    \right] \right. \\
    && \left. +
    t'\varepsilon \left[ \varepsilon^*(\lambda_R),
    j_2(\lambda_2,\lambda_2'),j_1(\lambda_1,\lambda_1') ,k_1 \right] \right), 
  \end{array}
  \label{For3P1AmpFor}
\end{equation}
\begin{equation}
  \begin{array}{lcl}
    {\cal M}(^3P_2;\lambda_1,\lambda_2,\lambda_1',\lambda_2',\lambda_R) & = & 
    \frac{c_4e^2}{tt'} \left( k_1 \cdot k_2 j_{1\mu}(\lambda_1,\lambda_1')
    j_{2\nu}(\lambda_2,\lambda_2') \right. \\
    && + k_{1\mu}k_{2\nu}  j_1(\lambda_1,\lambda_1') \cdot 
    j_2(\lambda_2,\lambda_2')  \\
    && - k_{1\mu} j_{2\nu}(\lambda_2,\lambda_2')
    j_1(\lambda_1,\lambda_1') \cdot k_2 \\
    && \left. - k_{2\mu} j_{1\nu}(\lambda_1,\lambda_1')
    j_2(\lambda_2,\lambda_2') \cdot k_1 \right) 
    \varepsilon^{*\mu\nu}(\lambda_R),  
  \end{array}
  \label{For3P2Exp}
\end{equation}
\begin{equation}
  \begin{array}{lcl}
    {\cal M}(^1D_2;\lambda_1,\lambda_2,\lambda_1',\lambda_2',\lambda_R) & = & 
    \frac{c_5e^2}{tt'}\varepsilon^{*\mu\nu}(\lambda_R)
    k_{1\mu} k_{2\nu} \\ &&
    \varepsilon \left[ k_1,k_2,j_1(\lambda_1,\lambda_1'),
    j_2(\lambda_2,\lambda_2') \right],
  \end{array}    
  \label{For1D2TotMatEl}
\end{equation}
where
\begin{equation}
  c_1=g_0, \hspace{0.4cm}
  c_2=4g_1/W, \hspace{0.4cm}
  c_3=2\sqrt{6} g_1, \hspace{0.4cm}
  c_4=4\sqrt{3}Wg_1, \hspace{0.4cm}
  c_5=8\sqrt{30}g_2.
\end{equation}
Here we have introduced
\begin{equation}
  g_i=\frac{16e_q^2|{\cal R}^{(i)}(0)|\alpha}{(s+s'+u+u'-8m_e^2)^{i+1}}
      \sqrt{\frac{3\pi}{W}},
\end{equation}
with ${\cal R}^{(i)}(0)$ the (derivates of the) radial part of the wave 
function in the origin and $e_q$ the
fractional quark charge. The actual values are given in I.

Replacing the currents $j_1$, $j_2$ by $\varepsilon_1$
and $\varepsilon_2$, the polarization vectors of a virtual photon, will give 
the matrix element of the two-photon reaction, i.e.\ the second step in 
reaction (\ref{ReacForm}). That type of matrix element can 
be simpliflied
by choosing convenient axes to describe the photon polarization. It will be
the basis of the method of section 4.
In the expressions terms of the form $\varepsilon_i \cdot k_i$
or $j_i \cdot k_i$ are omitted, since they vanish in our calculations.


\section{Direct method of calculation}

In this section one way of evaluating the cross section 
(\ref{SigmaDef}) and the related density matrix will be discussed.
The motivation to deal with another method in the next section 
is that we want to obtain 
the numerical
results for observables
in two completely independent ways.

The method of this section is a 
$2\to 3$ particle cross section calculation without using the two-photon
intermediate step. For the differential cross section alone it would be 
sufficient to express $\sum|{\cal M}|^2$ in terms of the invariants
(\ref{ForSDef})-(\ref{ForTpDef}). In fact, those have been obtained and are
given elsewhere \cite{GaGaResPaper}. It makes a very fast numerical 
evaluation of the cross section possible. It is natural to perform this 
calculation in the laboratory frame as specified in equation (\ref{labpara}).

For the density matrices it is more practical to use a rest system of the 
resonance. Every experimental event can be rotated and boosted back to 
the resonance rest frame (RRF) by successively rotating the momentum 
$\vec{p}_R$ in the
direction of the $z$-axis and boosting it to rest. Thus the required Lorentz
transformation is
\begin{equation}
  {\cal L}_{LR} = {\cal L}_B{\cal L}_{\theta}
  {\cal L}_{\phi} =
   \left(
  \begin{array}{cccc}
     \gamma
     & -\gamma\beta\cos\phi\sin\theta
     & -\gamma \beta \sin \phi \sin \theta
     & -\gamma \beta \cos \theta
     \\
     0
     & \cos\phi\cos\theta
     & \sin\phi\cos\theta
     & -\sin \theta
     \\
     0
     & -\sin \phi
     & \cos \phi
        & 0
     \\
     -\gamma \beta
     & \gamma \cos\phi\sin\theta
     & \gamma \sin \phi \sin \theta   &
     \gamma \cos \theta
   \end{array} \right).
\end{equation}

In order to calculate the average density matrix in the RRF belonging to a 
cross section in a certain kinematical region in the laboratory system
one proceeds as follows. In the laboratory system one generates 
weighted events. For this we use the event generator GaGaRes, which
is described elsewhere \cite{GaGaResPaper}. The event is boosted to the
RRF with ${\cal L}_{LR}$. For the momenta in the RRF the amplitude
${\cal A}_{\lambda_i,\lambda}$ is calculated, where
$\lambda_i$ are the lepton helicites and $\lambda$ the resonance 
helicity. From these amplitudes the normalized density matrix
\begin{equation}
  \rho_{\lambda\lambda'}=\frac{\sum_{\lambda_i}
  {\cal A}_{\lambda_i,\lambda}{\cal A}_{\lambda_i,\lambda'}^*}
  {\sum_{\lambda_i,\lambda_R} 
  {\cal A}_{\lambda_i,\lambda_R}{\cal A}_{\lambda_i,\lambda_R}^*},
  \label{ForNorDensMat}
\end{equation}
is evaluated for this event. By performing the weighted sum of these individual
density matrices and dividing by the total 
weight one obtains the average density matrix
for the production of a resonance.
Of course, when events with weight $1$ are generated the
above procedure becomes simpler. Sometimes unnormalized density matrices
are needed, i.e.\ just the numerator of (\ref{ForNorDensMat}).

It is clear that a method to evaluate amplitudes is needed, preferably a fast
technique. For this the Weyl-van der Waerden (WvdW) spinor calculus is a
good tool, as has been shown in the literature \cite{Giele}. Once Dirac 
spinors, momenta and polarization vectors have been translated into WvdW
spinors, the matrix elements (\ref{For1S0TotMatEl})-(\ref{For1D2TotMatEl})
become spinorial quantities.

There now are two ways to continue. The amplitudes can be expressed in terms
of spinorial inner products. Evaluating numerically these products and their
combinations one obtains the amplitudes as a complex number.
Another way is to interpret the spinorial matrix element as the trace of a
string of $2\times 2$ matrices, which can be calculated numerically very
fast thanks to efficient matrix multiplication in FORTRAN90. The latter
method has been used in our numerical calculations, whereas the former turned
out to be useful as independent 
check. More about the spinorial translation and
procedure can be found in \cite{GaGaResPaper}. Here we only mention one 
particular detail, relevant for comparisons and cross checks.

Usually the spin-1 polarization vectors for a particle with momentum
\begin{equation}
  k^\mu=
  (k^0,|\vec{k}|\sin\theta \cos\phi,|\vec{k}|\sin\theta\sin\phi,|\vec{k}|
  \cos\theta)=
  (k^0,|\vec{k}|s c_\phi,|\vec{k}| s s_\phi, |\vec{k}| c ),
\end{equation}
are chosen to be
\begin{equation}
  \begin{array}{l}
    \varep_{\pm}^\mu=\frac{1}{\sqrt{2}}\left(0,\mp c c_\phi +i s_\phi,
    \mp  cs_\phi-ic_\phi,\pm s\right), \\
    \varep_{0}^\mu=\frac{k^0}{m}\left(\frac{|\vec{k}|}{k_0},s c_\phi,
    s s_\phi,c\right),
  \end{array} 
  \label{ForPolVecStan}
\end{equation}
for helicities $\pm 1$ and $0$. The spin-2 polarization tensors then usually
are taken as
\begin{equation}
  \begin{array}{lcl}
    \varepsilon^{\mu\nu}(\pm 2) & = & \varepsilon^{\mu}(\pm 1)
    \varepsilon^{\nu}(\pm 1), \\
    \varepsilon^{\mu\nu}(\pm 1) & = & \frac{1}{\sqrt{2}} \left(
    \varepsilon^{\mu}(\pm 1)\varepsilon^{\nu}(0) +
    \varepsilon^{\mu}(0)\varepsilon^{\nu}(\pm 1) \right), \\
    \varepsilon^{\mu\nu}(0) & = & \frac{1}{\sqrt{6}} \left(
    \varepsilon^{\mu}(+1)\varepsilon^{\nu}(-1)+2\varepsilon^{\mu}(0)
    \varepsilon^{\nu}(0)+\varepsilon^{\mu}(-1)\varepsilon^{\nu}(1)
    \right).
  \end{array}
  \label{CGDecompA}
\end{equation}
Of course this follows a certain 
convention and other choices may be made as can be seen in the literature.
In the implementation of the WvdW formalism the conventions
of \cite{Ditmaier} have been used. Unfortunately the polarization vectors and
tensors in this reference are different
\begin{equation}
  \begin{array}{lcl}
    \varepsilon_\pm^\mu(k)&=&\frac{e^{\mp i \phi}}{\sqrt{2}}
    (0,-c c_\phi \pm i s_\phi,
    - c s_\phi \mp i c_\phi , s),
  \\ 
    \varepsilon_0^\mu & = & \frac{k^0}{m}\left(
    \frac{|\vec{k}|}{k^0},s c_\phi , ss_\phi,
    c \right),
  \end{array}
  \label{ForPolVecDit}
\end{equation}

\begin{equation}
  \begin{array}{lcl}
    \varepsilon^{\mu\nu}(\pm 2) & = & \varepsilon^\mu(\pm 1) 
    \varepsilon^\nu(\pm 1),
  \\ 
    \varepsilon^{\mu\nu}(\pm 1) & = &
    \frac{\pm 1}{\sqrt{2}}
    \left(
     \varepsilon^\mu(\pm 1) 
     \varepsilon^\nu(0)+
     \varepsilon^\mu(0) 
     \varepsilon^\nu(\pm 1)
    \right),
  \\
    \varepsilon^{\mu\nu}(0) & = &
    \frac{1}{\sqrt{6}}
    \left(-
     \varepsilon^\mu(+1) 
     \varepsilon^\nu(-1)+2
     \varepsilon^\mu(0) 
     \varepsilon^\nu(0)-
     \varepsilon^\mu(-1) 
     \varepsilon^\nu(+1)
    \right).
  \end{array}
  \label{ForDitSpin2Def}
\end{equation}
When
we label the sets of polarization vectors and tensors of equations
(\ref{ForPolVecStan}) and (\ref{CGDecompA})
with the index $A$
and the polarization vectors and corresponding tensors in 
equations (\ref{ForPolVecDit}) and (\ref{ForDitSpin2Def})
with an index $B$, 
the two sets are related by
\begin{equation}
  \begin{array}{c}
    \varep_{B,\pm}^\mu=\pm e^{\mp i \phi} \varep_{A,\pm}^\mu\hspace{1.0cm}
    \varep_{B,0}^\mu=\varep_{A,0}^\mu,
    \\
    \varepsilon^{\mu\nu}_B(0)  =  \varepsilon^{\mu\nu}_A(0), \hspace{1.0cm}
    \varepsilon^{\mu\nu}_B(\pm 1)  =   e^{\mp i \phi}
    \varepsilon^{\mu\nu}_A(\pm 1), \hspace{1.0cm}
    \varepsilon^{\mu\nu}_B(\pm 2)  =  e^{\mp 2 i \phi}
    \varepsilon^{\mu\nu}_A(\pm 2).
  \end{array}
  \label{ForPolRels}
\end{equation}
These formulae give the relations between density matrices in the two
conventions. When an event has a density matrix $\rho_A$, evaluated with 
the set $A$, then $\rho_B$ can be obtained by multiplying each matrix element
with a certain phase factor. In a compact notation these 
factors are summarized as follows
\begin{equation}
  \rho_{\lambda\lambda',B}=e^{i\Delta \phi_{\lambda\lambda'}}
  \rho_{\lambda\lambda',A},
  \label{ForDPhiRel}
\end{equation}
with for the spin-1 density matrix elements
\begin{equation}
  \Delta \phi_{\lambda\lambda'}
  =
  \left(
  \begin{array}{ccccc}
    0            && \phi        && 2\phi+\pi \\
    -\phi        && 0           && \phi+\pi \\
    -(2\phi+\pi) && -(\phi+\pi) && 0
  \end{array}
  \right).
\end{equation}
For the spin-2 density matrix elements we find analogously the set of
phase factors
\begin{equation}
  \Delta \phi_{\lambda\lambda'}
  =
  \left(
  \begin{array}{ccccc}
    0     & \phi & 2\phi & 3\phi & 4\phi \\
    -\phi & 0    & \phi & 2\phi & 3\phi  \\
    -2\phi & -\phi & 0  & \phi & 2\phi   \\
    -3\phi &  -2\phi & -\phi & 0  & \phi \\
    -4\phi & -3\phi &  -2\phi & -\phi & 0 
  \end{array}
  \right).
\end{equation}


\section{Extended BGMS formalism}
A long time ago in a classic paper \cite{BGMS} it was 
advocated to write the cross section of 
reaction (\ref{ReacForm}) as a two-step process. First
virtual photons characterized by a density matrix $\rho_i$
are created which in a second step produce the resonance $R$.
In this BGMS formalism the cross section for resonance production
(\ref{SigmaDef}) takes in first instance the form
\begin{equation}
  \mbox{d}\sigma=\frac{\alpha^2}{Q_1^2Q_2^2}\rho_1^{\mu\mu'}\rho_2^{\nu\nu'}
  \delta^{(4)}(k_1+k_2-P)
  \frac
  {\sum M_{\mu\nu}M_{\mu'\nu'}^*}
  {\sqrt{(p_1\cdot p_2)^2-m_e^4}} 
  \frac{\mbox{d}^3\vec{p'}_1}{2E_1'}
  \frac{\mbox{d}^3\vec{p'}_2}{2E_2'}
  \mbox{d}\Gamma,
  \label{ForSigTwo}
\end{equation}
where
\begin{equation}
  \rho_i^{\mu\mu'} = \frac{1}{2Q_i^2}\sum j^\mu j^{*\mu'} =
   \frac{2}{Q_i^2}\left(p_i^\alpha p_i'^\beta
  +p_i'^\alpha p_i^\beta
  -\frac{1}{2}Q_i^2g^{\alpha\beta}\right).
\end{equation}

The index summation in equation
(\ref{ForSigTwo}) is now replaced by a summation over photon helicities
$\lambda_i$ and $\lambda_i'$ with values $\pm 1$, $0$.
Introducing the quantity
\begin{equation}
  M_{\lambda_1'\lambda_2'\lambda_1\lambda_2}=\frac{1}{2}
  (2\pi)^4 \int \delta^4(k_1+k_2-p_R) \sum_{\lambda_R} 
  M_{\lambda_1'\lambda_2',\lambda_R}^*
  M_{\lambda_1\lambda_2,\lambda_R}
  d \Gamma,
  \label{ForMllll}
\end{equation}
the cross section can be written as 
\begin{equation}
  d\sigma = \frac{\alpha^2}{32\pi^4Q_1^2Q_2^2}
  \frac{1}{\sqrt{(p_1\cdot p_2)^2-m_e^4}}
  \sum
  \rho_1^{\lambda_1\lambda_1'}
  \rho_2^{\lambda_2\lambda_2'}
  M_{\lambda_1'\lambda_2'\lambda_1\lambda_2}
  \frac{d\vec{p}_1'}{E_1'}
  \frac{d\vec{p}_2'}{E_2'}.
\end{equation}
It should be noted that $M_{\lambda_1'\lambda_2'\lambda_1\lambda_2}$
is evaluated in the two-photon rest system, the BGMS frame,
where the direction of the
photon originating from the positron is taken as $z$-axis. The polarization
vectors of both photons now become very simple as can be seen
from (\ref{ForPolVecStan}). Also $\lambda_R$ is completely fixed by
\begin{equation}
  \lambda_R=\lambda_1-\lambda_2,
\end{equation}
and the summation in (\ref{ForMllll}) consists of only one term.

The expression for the unnormalized density matrix of an event is 
obtained from the cross section formula when one introduces another
combination $\rho\rho M$.
\begin{equation}
  \Sigma_{\lambda\lambda'}=\sum_{\begin{array}{c} 
                                  \lambda_1,\lambda_2,\lambda_1',\lambda_2' \\
                                  \lambda=\lambda_1-\lambda_2 \\
                                  \lambda'=\lambda_1'-\lambda_2' 
                                 \end{array}}
  \rho_1^{\lambda_1\lambda_1'}\rho_2^{\lambda_2\lambda_2'}
  M_{{\lambda_1}'{\lambda_2}'\lambda_1\lambda_2}.
\end{equation}
It should be noted that for a specific $\lambda$ only 
specific $\lambda_1$, $\lambda_2$ combinations can contribute.
Using the explicit form of the photon density matrices
\cite{BGMS}
where only a restricted number of elements is independent and
using 
\begin{equation}
  M_{\lambda_1,\lambda_2,\lambda_R}=\eta_R M_{-\lambda_1,-\lambda_2,-\lambda_R}
  \label{ForEtaRDef}
\end{equation}
where $\eta_R = 1 (-1)$ for the ``normal'' (``abnormal'') $J^P$ series, one
obtains expressions for $\Sigma_{\lambda\lambda'}$ in terms of 
$M_{{\lambda_1}'{\lambda_2}'\lambda_1\lambda_2}$.
For every expression also new quantities are introduced, generalizations
of the quantities $\sigma_{AB}$ and $\tau_{AB}$ of \cite{BGMS}
with the help of
\begin{equation}
  X_{\gamma\gamma}=(k_1\cdot k_2)^2-k_1^2k_2^2. 
\end{equation}
The 
results are 
\begin{equation}
\Sigma_{2+2+} = \Sigma_{2-2-} =\repp\rtpp M_{+-+-}=
  4\sqrt{X_{\gamma\gamma}}\repp\rtpp\sigma_{TT}^B, 
 \label{ForSig2p2p}
\end{equation}

\renewcommand{\arraystretch}{1.2}
\begin{equation}
  \begin{array}{ll}
    \Sigma_{++}   = \Sigma_{--} & =  \repp \rtnn M_{+0+0} +
    \renn\rtpp M_{0+0+}
    -2\repn\rtpn  
    M_{0-+0} 
    \cos(\tilde{\phi})
  \\ &
    = 2\sqrt{X_{\gamma\gamma}}\left(
    \repp \rtnn \sigma_{TS} +
    \renn\rtpp \sigma_{ST}
    -4\repn\rtpn \cos(\tilde{\phi})\tau_{TS}^B
    \right),
  \end{array}
\end{equation}


\renewcommand{\arraystretch}{1.2}

\begin{equation}
  \begin{array}{lcl}
    \Sigma_{00} & = &2 \repp\rtpp M_{++++} + \renn \rtnn M_{0000} 
                  +2\repm\rtpm \cos(2\tilde{\phi}) M_{--++}  
  \\
                  &&-4\repn\rtpn\cos(\tilde{\phi}) 
                  M_{00++}
  \\
    & = & 
    2\sqrt{X_{\gamma\gamma}}\left(
    4 \repp\rtpp \sigma_{TT}^A + \renn \rtnn \sigma_{SS} 
                  +2\repm\rtpm \cos(2\tilde{\phi}) \tau_{TT}   \right.
  \\
    && \left.
                  -8\repn\rtpn\cos(\tilde{\phi}) \tau_{TS}^A \right).
  \end{array}
  \label{signnm}
\end{equation}
The various $\sigma$ and $\tau$ definitions follow from 
(\ref{ForSig2p2p})-(\ref{signnm}). In BGMS the following combinations are
used
\begin{equation} 
  \stt \equiv \stt^A+\stt^B = \frac{1}{4\sqrt{X_{\gamma\gamma}}} 
  ( M_{++++} + M_{+-+-}),
\end{equation}
\begin{equation} 
  \tts \equiv \tau_{TS}^A+\tau_{TS}^B 
  = \frac{1}{4\sqrt{X_{\gamma\gamma}}} 
   (
   M_{++00} + M_{-00+}).
\end{equation}
In the formulae $\tilde{\phi}$ is the angle between the two lepton scattering 
planes in the BGMS frame. 

For the off-diagonal elements we 
do the same. The results are
summarized below
\begin{equation}
  \Sigma_{+-} = \etipe \left( 2 \repn \rtpn \emip 
  M_{0++0} 
  -\repm\rtnn M_{-0+0} - \renn\rtpm\emtip M_{0+0-} \right),
\end{equation}
\begin{equation}
  \begin{array}{lcl}
    \Sigma_{+0} & = & i\eipe \left( \repp \rtpn \emip M_{+++0}
    -\repn\rtnn M_{00+0}  \right. 
    \\
    &&
    + \repm\rtpn M_{--+0}\eip
    -\repn\rtpm\emtip M_{--0+}
    \\ && \left.
    +\renn\rtpn\emip M_{000+}
    -\repn\rtpp M_{++0+}
    \right).
  \end{array}
\end{equation}
For the spin-2 resonances one has additionally
\begin{equation}
  \Sigma_{2+1+}=i\eipe\left( \repp\rtpn \emip M_{+0+-}
  -\repn\rtpp M_{0-+-}\right),
\end{equation}
\begin{equation}
  \begin{array}{lcl}
    \Sigma_{2+0}&=&-\etipe\left( \repp\rtpm\emtip M_{+++-}  
    -\repn\rtpn\emip M_{00+-} \right. 
  \\ &&
    \left.
    +\repm\rtpp M_{--+-} \right),
  \end{array}
\end{equation}
\begin{equation}
  \Sigma_{2+1-}=
  ie^{3i\tilde{\phi}_1} \left(
  -\repm\rtpn\emip M_{-0+-}+\repn\rtpm\emtip M_{0++-}
  \right),
\end{equation}
\begin{equation}
  \Sigma_{2+2-}=
  e^{4i\tilde{\phi}_1} \repm\rtpm\emtp M_{-++-}.
\end{equation}

Besides the $\tilde \phi$ dependence, for the off-diagonal 
elements there is also an overall $\tilde{\phi}_1$ dependence,
 $\tilde{\phi}_1$ being the azimuthal angle of the incoming 
positron in the BGMS frame.

In analogy with the $\sigma$ and the $\tau$ terms we can introduce
\begin{equation}
  \chi_{+0}=\frac{i}{2\sqrt{X_{\gamma\gamma}}} M_{+++0},
  \hspace{1.0cm}
  \chi_{0+}=\frac{i}{2\sqrt{X_{\gamma\gamma}}} M_{++0+},
\end{equation} 
\begin{equation}
  \xi_{+0} =\frac{i}{2\sqrt{X_{\gamma\gamma}}} M_{00+0},
  \hspace{1.0cm}
  \xi_{0+} =\frac{i}{2\sqrt{X_{\gamma\gamma}}} M_{000+},
\end{equation}
\begin{equation}
  \zeta_{++} =\frac{1}{2\sqrt{X_{\gamma\gamma}}} M_{+++-},
  \hspace{1.0cm}
  \zeta_{00}=\frac{1}{2\sqrt{X_{\gamma\gamma}}} M_{00+-},
\end{equation}
\begin{equation}
  \zeta_{+0}=\frac{i}{2\sqrt{X_{\gamma\gamma}}} M_{+0+-},
  \hspace{1.0cm}
  \zeta_{0+}=\frac{i}{2\sqrt{X_{\gamma\gamma}}} M_{0++-}.
\end{equation}
The $\zeta$ functions are only non-vanishing for spin-2 resonances.
For all above functions $\sigma_{AB}$ , $\ldots$, $\zeta_{AB}$ holds: when
$\zeta_{AB}\sim M_{\lambda_1'\lambda_2'\lambda_1\lambda_2}$ then
$\zeta_{BA}\sim M_{\lambda_2'\lambda_1'\lambda_2\lambda_1}$.


The off-diagonal elements can now be written as
\begin{equation}
  \Sigma_{+-} = \etipe\sqrt{X_{\gamma\gamma}} 
  \left( 4 \repn \rtpn \emip \eta_R \tau_{TS}^B
  -\repm\rtnn\eta_R\sigma_{TS} - \renn\rtpm\emtip\eta_R\sigma_{ST} \right),
  \label{tsigpm}
\end{equation}
\begin{equation}
  \begin{array}{lcl}
    \Sigma_{+0} & = & \eipe \sqrt{X_{\gamma\gamma}}
    \left( \repp\rtpn \emip \chi_{+0}
    -\repn\rtnn\xi_{+0} + \repm\rtpn\eta_R\eip\chi_{+0} \right.
    \\
    && \left. -\repn\rtpm\emtip\eta_R\chi_{0+}
    + \renn\rtpn\emip\xi_{0+}-\repn\rtpp\chi_{0+} \right),
  \end{array}
  \label{tsigpn}
\end{equation}
\begin{equation}
  \Sigma_{2+1+}=\eipe\sqrt{X_{\gamma\gamma}}\left(\repp\rtpn\emip \zeta_{+0}
  -\repn\rtpp\eta_R\zeta_{0+}\right),
  \label{tsigtpep}
\end{equation}
\begin{equation}
  \Sigma_{2+0}=-\etipe\left( \repp \rtpm \emtip \zeta_{++}
  -\repn\rtpn\emip\zeta_{00}+\repm\rtpp\eta_R\zeta_{++}\right),
  \label{tsigtpn}
\end{equation}
\begin{equation}
  \Sigma_{2+1-}=\e3ipe\sqrt{X_{\gamma\gamma}}\left( -\repm\rtpn \emip\eta_R\zeta_{+0}
  +\repn\rtpm\emtip\zeta_{0+}\right),
  \label{tsigtpem}
\end{equation}
\begin{equation}
  \Sigma_{2+2-} = \efipe \emtip \sqrt{X_{\gamma\gamma}}\repm \rtpm \eta_R \stt^B.
  \label{tsigtptm}
\end{equation}
In these formulae $\eta_R$ is the phase factor 
arising in equation (\ref{ForEtaRDef}).
The $\xi$ functions
and $\zeta_{00}$
vanish for the $^3P_1$ resonance ($M_{00\lambda\lambda'}$ and
$M_{\lambda\lambda'00} $ vanish).

The other off-diagonal elements can be obtained by using that the 
density matrix is Hermitian
\begin{equation}
  \Sigma_{\lambda\lambda'}=\Sigma_{\lambda'\lambda}^*,
\end{equation}
and by using that for the polarization vectors in the BGMS formalism
the following identity holds
\begin{equation} 
  \Sigma_{\lambda\lambda'}=(-1)^{\lambda+\lambda'}\Sigma_{-\lambda'-\lambda}.
\end{equation} 

The unnormalized density matrix after integration over the phase space
of the outgoing leptons, denoted by
$\Sigma^{{int}}$,
is then given by
\begin{equation}
  \Sigma_{\lambda\lambda'}^{{int}}=\int\frac{\alpha^2}{32\pi^4Q_1^2Q_2^2}
  \frac{1}{\sqrt{(p_1\cdot p_2)^2-m_e^4}}
  \Sigma_{\lambda\lambda'}
  \frac{\mbox{d}^3 \vec{p'}_1}{E_1'}\frac{\mbox{d}^3 \vec{p'}_2}{E_2'}.
\end{equation}
The trace should equal the total cross section. When we use the expressions
for the diagonal elements we indeed obtain the BGMS expression
\renewcommand{\arraystretch}{1.4}
\begin{equation}
  \begin{array}{lcl}
    \mbox{d}\sigma
    & = & \frac{\alpha^2}{16\pi^4Q_1^2Q_2^2}
    \sqrt{\frac{(k_1\cdot k_2)^2-k_1^2k_2^2}{(p_1\cdot p_2)^2-m_e^4}}
    \left[ 4\rho_1^{++}\rho_2^{++}\sigma_{TT} \right. 
     + 2|\rho_1^{+-}\rho_2^{+-}|\tau_{TT}\cos(2\tilde{\phi})
     \\ &&
     +
     2\rho_1^{++}\rho_2^{00}\sigma_{ST} +
     2\rho_1^{00}\rho_2^{++}\sigma_{TS} 
     +\rho_1^{00}\rho_2^{00}\sigma_{SS}
     \\ && \left.
     -8|\rho_1^{+0}\rho_2^{+0}|\cos(\tilde{\phi})\tau_{TS} \right]
     \frac{\mbox{d}^3 \vec{p'}_1}{E_1'}\frac{\mbox{d}^3 \vec{p'}_2}{E_2'},
  \end{array}
  \label{ForBGMS}
\end{equation}
The explicit expressions for quantities $\sigma_{AB}$, $\tau_{AB}$,
 $\chi_{AB}$,  $\xi_{AB}$ and  $\zeta_{AB}$
follow from the amplitudes 
${M}_{\lambda_1,\lambda_2,\lambda_R}$ 
as given in table 1 of I.

In order to have some analytic results we repeat and extend form factor 
definitions of I and give tables with their forms. Thus we introduce
\begin{equation}
  \sigma_{AB}=\delta(P^2-M^2)8\pi^2\frac{(2J+1)\Gamma_{\gamma\gamma}(J^P)}
  {M}f_{AB}(J^P),
  \label{ForFFDef1}
\end{equation}
\begin{equation}
  \tau_{AB}=\delta(P^2-M^2)8\pi^2\frac{(2J+1)\Gamma_{\gamma\gamma}(J^P)}
  {M}g_{AB}(J^P).
  \label{ForFFDef2}
\end{equation}
 \begin{equation}
  \chi_{AB}=\delta(P^2-M^2)8\pi^2\frac{(2J+1)\Gamma_{\gamma\gamma}(J^P)}
  {M}k_{AB}(J^P),
  \label{ForFFDefChi}
\end{equation}
\begin{equation}
  \xi_{AB}=\delta(P^2-M^2)8\pi^2\frac{(2J+1)\Gamma_{\gamma\gamma}(J^P)}
  {M}m_{AB}(J^P),
  \label{ForFFDefXi}
\end{equation}
\begin{equation}
  \zeta_{AB}=\delta(P^2-M^2)8\pi^2\frac{(2J+1)\Gamma_{\gamma\gamma}(J^P)}
  {M}n_{AB}(J^P).
  \label{ForFFDefZeta}
\end{equation}
The form factors are given in tables \ref{TabFFs1}
and \ref{TabFFs2}, where the quantities
$\kappa$, $\nu$ and $X$ are defined as
\begin{equation}
  \kappa = \frac{M^2}{2\sqrt{X}},
\end{equation}
\begin{equation}
  \nu  =  k_1 \cdot k_2, 
\end{equation}
\begin{equation}
  X= X_{\gamma\gamma}    =  \nu^2-k_1^2k_2^2.
\end{equation}

\renewcommand{\arraystretch}{1.8}
\begin{table}[t]
  \[
    \begin{array}{|c|c|c|c|c|} \hline
      J^P & f_{TT}^A & f_{TT}^B & f_{TS} & f_{SS} 
      \\ \hline
      0^- & \kappa \frac{X}{\nu^2} & 0 & 0 & 0 
      \\ \hline
      0^+ & \kappa \left( \frac{X+\nu M^2}{3\nu^2}\right) &0 & 0 &
      2 \kappa \left(\frac{M^2\sqrt{Q_1^2Q_2^2}}{3\nu^2}\right)^2
      \\ \hline
      1^+ & \kappa \left(\frac{Q_1^2-Q_2^2}{2\nu}\right)^2 & 0 & 2\kappa
      \frac{M^2}{2\nu}\frac{Q_2^2}{2\nu}\left(\frac{\nu+Q_1^2}{\nu}\right)^2
      & 0 
      \\ \hline
      2^+ &  \kappa \left( \frac{M^2}{2\nu}\right)^2 & 
      \kappa \left( \frac{M^2}{2\nu}\right)^2
      \frac{[2Q_1^2Q_2^2-\nu(Q_1^2+Q_2^2)]^2}{6M^4\nu^2} 
      & \kappa \frac{M^2{Q_2^2}(\nu-Q_1^2)^2}{4\nu^4} &
      \kappa \frac{M^4Q_1^2Q_2^2}{3\nu^4} 
      \\ \hline
      2^- & \kappa \left[ \frac{X}{\nu^2} \right]^3 & 0 & 0 & 0 
      \\ \hline
    \end{array}
  \]
  \[
    \begin{array}{|c|c|c|} \hline
      J^P & g_{TT} & g_{TS} 
      \\ \hline
      0^- & -2\kappa \frac{X}{\nu^2} & 0 
      \\ \hline
      0^+ & 2 \kappa \left( \frac{X+\nu M^2}{3\nu^2}\right) & 
      \kappa \left( \frac{M}{3\nu^2}\right)^2 Q_1 Q_2 (X+\nu M^2)
      \\ \hline
      1^+ & -2\kappa \left(\frac{Q_1^2-Q_2^2}{2\nu}\right)^2  & 
      \kappa\left( \frac{M}{2\nu^2} \right)^2
      Q_1Q_2(\nu+Q_1^2)(\nu+Q_2^2)
      \\ \hline
      2^+ & \frac{\kappa}{12} \left( \frac{2Q_1^2Q_2^2-\nu(Q_1^2+Q_2^2)}{\nu^2}
      \right)^2  & \kappa \left( \frac{M}{\nu^2}\right)^2 \frac{Q_1Q_2}
      {8}\left( \frac{2}{3}(2Q_1^2Q_2^2-\nu(Q_1^2+Q_2^2))
      +(\nu-Q_1^2)(\nu-Q_2^2)\right)
      \\ \hline
      2^- &  -2\kappa \left[ \frac{X}{\nu^2} \right]^3 & 0
      \\ \hline
    \end{array}
  \]
  \caption{{\it The form factors required for the cross section $f_{AB}$, $g_{AB}$, 
  ($A,B=T,S$).}}
  \label{TabFFs1}
\end{table}
\begin{table}
  \[
    \begin{array}{|c|c|c|} \hline
       J^P & k_{+0} & m_{+0}  \\ \hline
       1^+ & -\kappa\frac{M}{2}\frac{Q_2}{\nu^3}(\nu+Q_1^2)(Q_1^2-Q_2^2) &
             0
    \\ \hline
       2^+ & -\kappa \frac{MQ_2}{4\sqrt{3}}\frac{(\nu-Q_1^2)}{\nu^4}
       \left[ 2Q_1^2Q_2^2-\nu(Q_1^2+Q_2^2)\right] 
       & 
       \kappa \frac{M^3Q_1Q_2^2}{2\sqrt{3}\nu^4}(\nu-Q_1^2)
    \\ \hline
       2^- & 0 & 0 
    \\ \hline
    \end{array}
  \]
  \[
    \begin{array}{|c|c|c|c|} \hline
        J^P & n_{++} & n_{+0} & n_{00}  \\ \hline
        2^+ & \kappa \frac{1}{2\sqrt{6}} \frac{M^2}{\nu^3}
              \left[ 2Q_1^2Q_2^2-\nu(Q_1^2+Q_2^2)\right]
        &
           \kappa \frac{\sqrt{2}}{4}\frac{M^3Q_2}{\nu^3}(\nu-Q_1^2)
        &
           -\kappa \frac{M^4Q_1Q_2}{\sqrt{6}\nu^3}
    \\ \hline
        2^- & 0 & 0 & 0
    \\ \hline
    \end{array}
  \]
  \caption{{\it The additional form factors required for the off-diagonal elements.
  $k_{AB}$, $m_{AB}$ and $n_{AB}$ ($A,B=+,0$).}}
  \label{TabFFs2}
\end{table}

It should be stressed that for every event the matrix $\Sigma$
can now 
be evaluated. Although it is calculated in the two-photon rest frame, 
which is 
a frame where the resonance is at rest, it is not the RRF of the previous
section. For every event one has to relate the axes by a rotation
which gives rise to a transformation of the density matrix.
The transformation matrices are given in appendix A.
When one likes to compare the average density matrix to the one of the previous
section, one should also perform the
change (\ref{ForDPhiRel}) for every event.

With the above procedure 
we extend the existing BGMS cross section Monte Carlo program 
Galuga \cite{Galuga} to a program which can calculate the 
density matrix $\Sigma$. We then evaluate
for every event generated by this 
program the 
density matrix 
$\Sigma$ which is then transformed to the experimentally relevant density
matrix of section 3.
In this
way Galuga can evaluate 
the wanted density matrix. 
Moreover, we thus have an independent check on the GaGaRes 
evaluation.


\section{Example of a decay model and density matrix}

In general the produced resonance $R$ will decay into some final state 
$X$, so that the reaction is
\begin{equation}
  e^+e^- \to e^+e^- R \to e^+e^-X.
\end{equation}
The total amplitude for this process can be written as
\begin{equation}
  {\cal M} = \sum_{\lambda_R} \xi(P,M) {\cal A}_{\lambda_R}
  {\cal D}_{\lambda_R}.
\end{equation}
${\cal A}_{\lambda_R}$ describes the two-photon production of a resonance
with helicity $\lambda_R$. ${\cal D}_{\lambda_R}$ describes the decay of the
resonance with helicity $\lambda_R$ into the final state $X$. The factor
$\xi(P,M)$ represents the propagator of the resonance and numerical
factors. The total matrix element still depends on the external four-momenta.
No implicit integrations over the outgoing momenta have been carried out
at this stage.
  The square of the total matrix element is given by
\begin{equation}
  \sum |{\cal M}|^2 = |\xi(P,M)|^2\sum_{\lambda_R,\lambda_R'}
  {\cal A}_{\lambda_R\lambda_R'}{\cal D}_{\lambda_R\lambda_R'}
  =  |\xi(P,M)|^2\mbox{Tr}\left({\cal A}{\cal D}^*\right).
  \label{ForTraceDensMatDef}
\end{equation}
The summation on the left hand side represents a summation over the 
helicities of the initial and final state particles. The quantities
$ {\cal A}_{\lambda_R\lambda_R'}$ and ${\cal D}_{\lambda_R\lambda_R'}$
are ${\cal A}_{\lambda_R}{\cal A}_{\lambda_R'}^*$ and
${\cal D}_{\lambda_R}{\cal D}_{\lambda_R'}^*$, summed over the helicities
of all particles but the resonance. These are the density matrices for the
production and decay of the resonance. For a spin-$J$ resonance the 
density matrices are formed by
$(2J+1)\times(2J+1)$-matrices. For actual calculations a choice of polarization
vectors and tensors has to be made. The results of this paper are obtained
in the conventions (\ref{ForPolVecDit}) and (\ref{ForDitSpin2Def}).

For $R$ we first 
take the spin-2 state $\chi_{c2}$ and consider 
the specific
decay mode
\begin{equation}
  \chi_{c2}(p_R) \to \gamma(k_1) J/\psi(k_2).
  \label{ForResDec}
\end{equation}
Later on also the possibility of the $J/\psi$ decaying into a lepton pair
will be discussed. The choice of this example is prompted by its
experimental relevance.
Since we expect that a similar $\chi_{c1}$ decay may cause a contamination
of the events (\ref{ForResDec}) in an actual experiment we shall  also discuss the $\chi_{c1}$ decay at the end 
of this section.

For the amplitude ${\cal M}$
of the decay of the spin-2 resonance we use the amplitude that
is given in equation (\ref{For3P2Exp})
\begin{equation}
  \begin{array}{lcl}
  {\cal M}(\lambda,\lambda_1,\lambda_2) & = & \tilde{c}_4\left[
  (k_1\cdot k_2) \varepsilon_1^{*\mu}(\lambda_1) \varepsilon_2^{*\nu}
  (\lambda_2) +
  (\varepsilon_1^*(\lambda_1) \cdot \varepsilon_2^*(\lambda_2)) k_1^\mu k_2^\nu
  \right.
  \\ &&
  \left.
  -(k_1 \cdot \varepsilon_2^*(\lambda_2)) \varepsilon_1^{*\mu}(\lambda_1) 
  k_2^\nu
  -(k_2 \cdot \varepsilon_1^*(\lambda_1)) \varepsilon_2^{*\mu}(\lambda_2) 
  k_1^\nu
  \right]\varepsilon_{\mu\nu}(\lambda)
  \\ & = & 
  \tilde{c}_4 F_1^{*\rho\mu}(\lambda_1)F_{2\rho\nu}^*(\lambda_2) \varepsilon_{\mu}^{\nu}
  (\lambda),
  \end{array}
  \label{ForMatExp3P2}
\end{equation}
where $F_i^{\alpha\beta}$ is the field strength
\begin{equation}
  F_i^{\alpha\beta}(\lambda_i)=k_i^\alpha\varepsilon_i^\beta(\lambda_i)-
  k_i^\beta\varepsilon_i^\alpha(\lambda_i).
\end{equation}
In these equations the index
$1$ refers to the photon  and $2$ refers to the
massive $J/\psi$. The $\lambda$'s refer to the helicity of the
associated particles. 
The parameter $\tilde{c}_4$ is not specified here, 
but its complex square will contain
the width for the decay of the spin-2 particle into these two 
spin-1 particles. It will be related to $c_4$.

The density matrix ${\cal D}$ 
for the decay of the resonance is then constructed
by
\begin{equation}
  {\cal D}_{\lambda\lambda'}=\sum_{\lambda_1,\lambda_2}
  {\cal M}(\lambda,\lambda_1,\lambda_2)
  {\cal M}^*(\lambda',\lambda_1,\lambda_2).
\end{equation}
In the evaluation of this density matrix we make use of the 
invariant spin summations for spin-1 particles. For the massless photon
this 
relation reads
\begin{equation}
  \sum_{\lambda_1} \varepsilon_{1,\alpha}(\lambda_1) 
  \varepsilon_{1,\beta}^*(\lambda_1)
  = - g_{\alpha\beta},
  \label{CompRelMassLess}
\end{equation}
whereas for the massive $J/\psi$ 
it reads
\begin{equation}
  \sum_{\lambda_2} \varepsilon_{2,\alpha}(\lambda_2)
  \varepsilon_{2,\beta}^*(\lambda_2)
  = - g_{\alpha\beta}  + \frac{k_{2,\alpha}k_{2,\beta}}{M^2},
  \label{CompRelMassive}
\end{equation}
where $M$ is 
its mass.
In fact, due to the gauge invariance of the field strength in equation
(\ref{ForMatExp3P2}) only the $g_{\alpha\beta}$ term in
equation (\ref{CompRelMassive}) contributes.
This leads to the following expression for the density matrix for the
decay of the resonance
\begin{equation}
  \begin{array}{lcl}
    {\cal D}_{\lambda\lambda'} & = & |\tilde{c}_4|^2\left[ (k_1\cdot k_2)
    \varepsilon^{\alpha\beta}(\lambda)\varepsilon_{\alpha\beta}^*(\lambda')
    \right.
    +2(k_1\cdot\varepsilon(\lambda)\cdot k_1) (k_1\cdot\varepsilon^*
    (\lambda')\cdot k_1)
    \\ &&
    \left.
    (k_2^2+4(k_1\cdot k_2))(k_1\cdot\varepsilon(\lambda)\cdot
    \varepsilon^*(\lambda')\cdot k_1)
    \right].
  \end{array}
\end{equation}
In this equation a polarization tensor in between two dots indicates that
one Lorentz index has to be contracted with the 
four-vector on the 
left side of the tensor whereas the other Lorentz index has to be contracted
with the 
four-vector on the right side of the tensor, i.e. 
\begin{equation}
  p \cdot A \cdot q \equiv A^{\mu\nu}p_\mu q_\nu.
\end{equation}
In the derivation
we have used that the final state photon is massless and 
 we have used 
the properties of the polarization tensor 
to replace
$k_2$ in contractions with 
it by $-k_1$.

The density matrix for the decay of the $\chi_{c2}$ is 
constructed in the rest frame of the resonance with the $z$ axis as
polarization axis and some $x$ and $y$ axes. The density matrix will depend on
the polar and azimuthal angles of the photon in this reference system.
The specific choice of the axes in the $\chi_{c2}$ rest system is at this
point arbitrary. In the applications they should be the same as for the 
$\chi_{c2}$ production density matrix.
The polarization tensor looks very simple and one
can use equations (\ref{ForPolVecDit}) and (\ref{ForDitSpin2Def})
to verify that the following useful relations hold
\begin{equation}
  \varepsilon^\mu(\lambda) \varepsilon_\mu^*(\lambda')=
  -\delta_{\lambda\lambda'}, 
  \hspace{1.0cm}
  \varepsilon^{\mu\nu}(\lambda) \varepsilon_{\mu\nu}^*(\lambda')\equiv
  \delta_{\lambda\lambda'}.
\end{equation}

In this frame the four-momentum of particle $1$ can be parametrized 
by
\begin{equation}
  k_1^\mu = |\vec{k}|(1,\sin\theta\cos\phi,\sin\theta\sin\phi,\cos\theta)
  =|\vec{k}|(1,s \cos\phi,s \sin \phi,c).
  \label{ForK1ParRRF}
\end{equation}

It can be shown that with these definitions the density matrix for the
decay of the spin-2 particle can be written in a compact form as the sum
of three matrices
\begin{equation}
  \begin{array}{lcl}
    {\cal D}_{\lambda\lambda'}&=&|\tilde{c}_4|^2\left[
    (k_1\cdot k_2)^2 \delta_{\lambda\lambda'}
    +2|\vec{k}|^4 v_\lambda v_{\lambda'}^* +
    (k_2^2+4(k_1\cdot k_2))|\vec{k}|^2 R_{\lambda\lambda'}
  \right] 
  \\ &=&
    |\tilde{c}_4|^2\left(\frac{M_R^2-M^2}{2}\right)^2\left[
    \delta_{\lambda\lambda'}
    +\frac{1}{2}\alpha^2v_{\lambda}v_{\lambda'}^*+(1+\alpha)R_{\lambda\lambda'}
    \right].
  \end{array}
  \label{ForDecMatTot}
\end{equation}
In this expression $\delta$ is the Kronecker symbol. The vector $v$ is 
given by
\begin{equation}
  v=(v_2,v_1,v_0,v_{-1},v_{-2})=
  (\frac{s^2}{2}\tfp,-sc\fp,\frac{1}{\sqrt{6}}(3c^2-1),
  sc\fm,\frac{s^2}{2}\tfm).
\end{equation}
The tensor $R$ is given by (the indices run from $2$ to $-2$)
\begin{equation}
  R_{\lambda\lambda'}=
  \left(
  \begin{array}{ccccc}
     -\frac{s^2}{2} & \frac{cs}{2}\fp & \frac{s^2}{2\sqrt{6}}\tfp & 0 & 0 \\
     \frac{cs}{2}\fm   & -\frac{1+c^2}{4} & \frac{sc}{2\sqrt{6}}\fp & 
     \frac{s^2}{4}\tfp & 0 \\
     \frac{s^2}{2\sqrt{6}}\tfm & \frac{sc}{2\sqrt{6}}\fm & -\frac{3c^2+1}{6} & 
     -\frac{sc}{2\sqrt{6}}\fp & \frac{s^2}{2\sqrt{6}}\tfp \\
     0 & \frac{s^2}{4}\tfm & -\frac{sc}{2\sqrt{6}}\fm & -\frac{1+c^2}{4} & 
     -\frac{cs}{2}\fp \\
     0 & 0 & \frac{s^2}{2\sqrt{6}}\tfm & -\frac{cs}{2}\fm & -\frac{s^2}{2}
  \end{array}
  \right).
\end{equation}
In the second line of equation (\ref{ForDecMatTot}) the expansion parameter
$\alpha$ has been introduced.
\begin{equation}
  \alpha=1-\frac{M^2}{M_R^2}.
\end{equation}

Others 
\cite{ChoWiseTrivedi} use a simplified expression for the matrix element,
based on an electric dipole transition
\begin{equation}
  {\cal M} = \tilde{c}_4 \left[
  (k_1\cdot k_2) \varepsilon_1^{*\mu}(\lambda_1)\varepsilon_2^{*\nu}(\lambda_2)
  -(k_2\cdot \varepsilon_1^*(\lambda_1))k_1^\mu\varepsilon_2^{*\nu}(\lambda_2) 
  \right]
  \varepsilon_{\mu\nu}(\lambda).
\end{equation}
This leads to the following density matrix
\begin{equation}
  \begin{array}{lcl}
    {\cal D}_{\lambda\lambda'} & = &|\tilde{c}_4|^2\left[
    (k_1\cdot k_2)^2 \delta_{\lambda\lambda'}
    -\frac{1}{M^2}(k_2^2+2(k_1\cdot k_2))|\vec{k}|^4 v_\lambda v_{\lambda'}^* +
    (k_2^2+2(k_1\cdot k_2)
    \right.
  \\ &&
    \left.
    -\frac{(k_1\cdot k_2)^2}{M^2})|\vec{k}|^2
    R_{\lambda\lambda'}
    \right]
  \\ &=&
    |\tilde{c}_4|^2\left(\frac{M_R^2-M^2}{2}\right)^2\left[
    \delta_{\lambda\lambda'}-\frac{\alpha^2}{4(1-\alpha)}v_\lambda 
    v_{\lambda'}^*
    +\left(1-\frac{\alpha^2}{4(1-\alpha)}\right)R_{\lambda\lambda'}\right].
  \end{array}
  \label{ForDecMatDP}
\end{equation}
In the limit of vanishing $\alpha$ the density matrices of 
equations (\ref{ForDecMatTot}) and 
(\ref{ForDecMatDP}) 
become equal. For the actual $\alpha$,
based on $M=3.09687 \ \mbox{GeV}$, $M_R=3.55618 \ \mbox{GeV}$ \cite{PDG} 
and therefore $\alpha\approx 0.242$,
there 
are some differences, e.g. for equation (\ref{ForDecMatTot})
\begin{equation}
    {\cal D}_{22}+{\cal D}_{-2-2}  \sim  1  +  1.570 \cdot c^2  +  
    0.0190 \cdot c^4,
    \label{ForForXXX}
\end{equation}
and for equation (\ref{ForDecMatDP})
\begin{equation}
    {\cal D}_{22}+{\cal D}_{-2-2} \sim 1  +  1.028 \cdot c^2  -  
    0.00956 \cdot c^4,
    \label{ForForYYY}
\end{equation}
whereas for $\alpha=0$ one would have
\begin{equation}
  {\cal D}_{22}+{\cal D}_{-2-2}  \sim  1 + c^2.
  \label{ForAlp0lim}
\end{equation}
These distributions can be compared to the experimentally observed decay 
distribution \cite{E760Chic2} of the $\chi_{c2}$ produced in $p\bar{p}$
collisions,
\begin{equation}
  \begin{array}{lclclcll}
    {\cal D}_{22}+{\cal D}_{-2-2} & \sim & 1 & + & 1.96 \cdot  
    c^2 & + & 0.0142 \cdot c^4,
    & 
  \end{array}
  \label{ForExpDis22}
\end{equation}
where the central values of the coefficients are taken and the errors are
omitted. 
In section 6 some numerical results for the distributions will be presented.


One could also take into account the possible decay of the massive $J/\psi$ 
into two leptons
\begin{equation}
  \chi_{c2}(k)\to \gamma(k_1) J/\psi(k_2) \to \gamma(k_1) 
  l^+(p_1) l^-(p_2).
\end{equation}
The matrix element for this reaction can be obtained from equation
(\ref{ForMatExp3P2}) by replacing the polarization vector of the 
$J/\psi$ by the lepton current
\begin{equation}
  J_{rs}^\mu = \bar{u}_r(p_2) \gamma^\mu v_s(p_1),
\end{equation}
where $r$ and $s$ denote the spins of the final state leptons.
In the following the lepton mass $m_l$ will be neglected, which is a reasonable
approximation for this decay.

In the construction of the density matrix for this decay one has to use
that for the lepton current one finds
\begin{equation}
  \sum_{r,s} J_{rs}^\mu J_{rs}^{*,\nu} = 4[p_1^\mu p_2^\nu + p_1^\nu p_2^\mu
  -\frac{1}{2}k_2^2 g^{\mu\nu}]=
  2 [ k_2^\mu k_2^\nu - l^\mu l^\nu - k_2^2 g^{\mu\nu}].
\end{equation}
In this equation the vector $l=p_1-p_2$ has been introduced. This has been
done as the term proportional to $k_2^\mu k_2^\nu$ will not contribute to
the density matrix.
For $l$ we use the following parametrization
\begin{equation}
  l^\mu=(l_0,l_x,l_y,l_z)
\end{equation}
in the above chosen reference frame, where (\ref{ForK1ParRRF}) holds.
The density matrix now reads
\begin{equation}
  \begin{array}{lcl}
    {\cal D}_{\lambda\lambda'} & = & |\tilde{c}|^2
    \left(
    \varepsilon^{\alpha\beta}(\lambda)\varepsilon_{\alpha\beta}^*(\lambda')
    (k_1\cdot k_2)^2 k_2^2 
    \right.
    \\ &&
    +(2k_2^2+l^2)(k_1\cdot\varepsilon(\lambda)\cdot k_1)
    (k_1\cdot\varepsilon^*(\lambda')\cdot k_1)
    \\ && 
    -(k_1\cdot k_2)[ (k_1\cdot\varepsilon(\lambda)\cdot k_1)
    (l\cdot\varepsilon^*(\lambda')\cdot l)+
    (l\cdot\varepsilon(\lambda)\cdot l)
    (k_1\cdot\varepsilon^*(\lambda')\cdot k_1) ]
    \\ &&
    +(2(k_1 \cdot k_2)+k_2^2 ) (k_1\cdot\varepsilon(\lambda)\cdot l)
    (k_1\cdot\varepsilon^*(\lambda')\cdot l) 
    \\ &&
    +  [(k_2^2)^2+4(k_1\cdot k_2)k_2^2+(k_1 \cdot l)^2] 
    (k_1\cdot \varepsilon(\lambda)
    \cdot \varepsilon^*(\lambda') \cdot k_1)
    \\ &&
    +(k_1\cdot k_2) (k_1 \cdot l)[ 
    (k_1\cdot \varepsilon(\lambda)
    \cdot \varepsilon^*(\lambda') \cdot l)
    + 
    (l\cdot \varepsilon(\lambda)
    \cdot \varepsilon^*(\lambda') \cdot k_1)] 
    \\ &&
    \left.
    +(k_1\cdot k_2)^2 (l\cdot \varepsilon(\lambda)
    \cdot \varepsilon^*(\lambda') \cdot l)
    \right).
  \end{array}
\end{equation}
In the derivation we have used that $(k_2 \cdot l)=0$
This density matrix can also be written in a more compact form
\begin{equation}
  \begin{array}{lcl}
    {\cal D}_{\lambda\lambda'} & = & 
    |\tilde{c}|^2
    \left(
    (k_1\cdot k_2)^2 k_2^2 \delta_{\lambda\lambda'}
    \right. 
    \\ &&
    + [2k_2^2+l^2]|\vec{k}|^4v_\lambda v_{\lambda'}^*
    -(k_1\cdot k_2)|\vec{k}|^2
    [v_\lambda x_{\lambda'}^*+x_\lambda v_{\lambda'}^*]
    \\ &&
    +(2(k_1\cdot k_2)+k_2^2)|\vec{k}|^2w_\lambda w_{\lambda'}^*
    \\ &&
    +[(k_2^2)^2+
    4(k_1\cdot k_2)k_2^2+(k_1\cdot l)^2]|\vec{k}|^2R_{\lambda\lambda'} 
    \\ &&
    \left.
    +(k_1\cdot k_2)(k_1 \cdot l)|\vec{k}|
    [S_{\lambda\lambda'}+S_{\lambda'\lambda}^*]
    +(k_1\cdot k_2)^2 T_{\lambda\lambda'} \right).
  \end{array}
\end{equation}
In this density matrix the coefficients can again be written as 
functions of $\alpha$
\begin{equation}
  \begin{array}{lcl}
    {\cal D}_{\lambda\lambda'} & = & 
    |\tilde{c}|^2
    \left(\frac{M_R^2-M^2}{2}\right)^2M^2\left(
    \delta_{\lambda\lambda'}+\frac{\alpha^2}{4}v_\lambda v_{\lambda'}^*
    -\frac{\alpha}{2M^2}(v_\lambda x_{\lambda'}^*
    +x_\lambda v_{\lambda'}^*)
    \right.
    \\ &&
    +\frac{1}{M^2}w_\lambda w_{\lambda'}^*
    +(1+\alpha+\frac{\alpha^2}{4(1-\alpha)} \cos^2 \theta^*)
    R_{\lambda\lambda'}
    \\ &&
    \left.
    \frac{\alpha}{2M\sqrt{1-\alpha}}\cos \theta^*(S_{\lambda\lambda'}
    +S_{\lambda'\lambda}^*)+\frac{1}{M^2}T_{\lambda\lambda'}\right),
  \end{array}
\end{equation}
where a convenient expression
\begin{equation}
  k_1 \cdot l = \frac{M_R^2-M^2}{2}\cos \theta^*
\end{equation}
has been used.
The angle  $\theta^*$ is the polar angle of the outgoing $l^+$ in the
rest system of the $J/\psi$ where the $z$ axis is given by
the direction of the boost from the RRF to the rest frame of the
$J/\psi$, i.e.\ opposite to the photon direction.
In the expressions some additional vectors and tensors have been 
introduced. 
\begin{equation}
  w=
  (\frac{sl_+}{2}\fp,-\frac{1}{2}[l_zs\fp+l_+c],
  -\frac{1}{\sqrt{6}}[s(l_x\cos\phi+l_y\sin\phi)-2cl_z],
  \frac{1}{2}[l_zs\fm+l_-c],\frac{sl_-}{2}\fm
  ),
\end{equation}
\begin{equation}
 x=(\frac{1}{2}l_+^2,-l_+l_z,-\frac{1}{\sqrt{6}}[l_x^2+l_y^2-2l_z^2],
 l_-l_z,\frac{1}{2}l_-^2),
\end{equation}
\begin{equation}
  \begin{array}{lcl}
    S_{\lambda\lambda'} & = &
    \left(
    \begin{array}{ccccc}
      -\frac{sl_-}{2}\fp & \frac{sl_z}{2}\fp & \frac{sl_+}{2\sqrt{6}}\fp 
      & 0 & 0
    \\
      \frac{cl_-}{2} & -\frac{1}{2}l_{sz} & 
      \frac{l_{s+}}{2\sqrt{6}} & \frac{sl_+}{4}\fp & 0  
    \\
      \frac{sl_-}{2\sqrt{6}}\fm & \frac{l_{cz}}{2\sqrt{6}} &
      -\frac{1}{6}[sl_{xy}+4cl_z]&
      - \frac{l_{cz}^*}{2\sqrt{6}} & 
      \frac{sl_+}{2\sqrt{6}}\fp
    \\
      0 & \frac{sl_-}{4}\fm & -\frac{l_{s+}^*}{2\sqrt{6}}&
      -\frac{1}{2}l_{sz}^* & -\frac{cl_+}{2} 
    \\
      0 & 0 &  \frac{sl_-}{2\sqrt{6}}\fm &
      -\frac{sl_z}{2}\fm & -\frac{sl_+}{2}\fm
    \end{array}
    \right),
  \end{array}
\end{equation}
\begin{equation}
  T_{\lambda\lambda'}=\left(
  \begin{array}{ccccc}
    -\frac{1}{2} (l_x^2+l_y^2) & \frac{l_+l_z}{2} & \frac{l_+^2}{2\sqrt{6}}
    & 0 & 0 
    \\
    \frac{l_-l_z}{2} & -\frac{1}{4}[|\vec{l}|^2+l_z^2] &
    \frac{l_+l_z}{2\sqrt{6}} & \frac{l_+^2}{4} & 0 
    \\
    \frac{l_-^2}{2\sqrt{6}} &  \frac{l_-l_z}{2\sqrt{6}} &
    -\frac{1}{6}[|\vec{l}|^2+3l_z^2] & - \frac{l_+l_z}{2\sqrt{6}} &
    \frac{l_+^2}{2\sqrt{6}}
    \\
    0 & \frac{l_-^2}{4} & - \frac{l_-l_z}{2\sqrt{6}} & 
    -\frac{1}{4}[|\vec{l}|^2+l_z^2] &
    -\frac{l_+l_z}{2} 
    \\
    0 & 0 & \frac{l_-^2}{2\sqrt{6}} & -\frac{l_-l_z}{2} & 
    -\frac{1}{2} (l_x^2+l_y^2)
  \end{array}
  \right).
\end{equation}
In the expressions the variables $l_+$ and $l_-$ have been introduced
\begin{equation}
  \begin{array}{lcl}
        l_+ & = & l_x + i l_y, \\
        l_- & = & l_x - i l_y.
  \end{array}
\end{equation}
In addition, we have also introduced
\begin{equation}
  l_{xy}=l_x\cos\phi+l_y\sin\phi,
\end{equation}
\begin{equation}
  l_{cz}=2cl_--sl_ze^{-i\phi},
\end{equation}
\begin{equation}
  l_{s+}=2sl_ze^{i\phi}-cl_+,
\end{equation}
\begin{equation}
  l_{sz}=\frac{sl_-}{2}e^{i\phi}+cl_z.
\end{equation}

The decay density matrix is now completely specified once the lepton
momenta in the $\chi_{c2}$ rest system are inserted in $l_\mu$.
It is sometimes convenient to parametrize $l_\mu$ in terms of
decay angles $\theta^*$ and $\phi^*$,
\begin{equation}
  s^* = \sin \theta^*, \hspace{1.0cm} 
  c^* = \cos \theta^*, \hspace{1.0cm}
  c_\phi^* = \cos \phi^*, \hspace{1.0cm}
  s_\phi^* = \sin \phi^*, 
\end{equation}
of the $l^+$ in the rest system of the $J/\psi$.

The components of the four-momentum $l$ in the RRF then become 
($c_\phi=\cos\phi$, $s_\phi=\sin\phi$, the angles of equation 
(\ref{ForK1ParRRF}))
\begin{equation}
  \begin{array}{lcl}
    l_0 & = & \frac{M_R^2-M^2}{2M_R}c^*, \\
    l_x & = & M(s^*(c^*cc_\phi+s^*s_\phi)-c^*sc_\phi),\\
    l_y & = & M(s^*(c^*cs_\phi+s^*c_\phi)-c^*ss_\phi),\\
    l_z & = & -M(s^*c_\phi^*s+c^*c).
  \end{array}
\end{equation}

Since the $\chi_{c2}$ will be predominantly produced in helicity $\pm 2$
states, we have a closer look at those density matrix elements.
In the limit of vanishing $\alpha$ the sum of density matrix elements 
${\cal D}_{-2-2}+{\cal D}_{22}$ reads
\begin{equation}
  \begin{array}{lcl}
    {\cal D}_{-2-2}+{\cal D}_{22} & = & 
    |\tilde{c}|^2\left(\frac{M_R^2-M^2}{2}\right)^2M^2\left(
    2+\frac{1}{M^2}(w_{-2}w_{-2}^*+w_{2}w_{2}^*) 
    \right.
  \\
    && 
    \left.
    + \frac{1}{M^2}(T_{-2-2}+T_{22})+R_{-2-2}+R_{22}
    \right).
  \end{array}
\end{equation}
Inserting the expressions for the vectors and the tensors
gives
\begin{equation}
  {\cal D}_{-2-2}+{\cal D}_{22} = 
  |\tilde{c}|^2\left(\frac{M_R^2-M^2}{2}\right)^2M^2
  (1+c^2)\left[1-\frac{l_+l_-}{2M^2}\right].
\end{equation}

Integrating over the angles associated to the outgoing leptons yields for the
sum ${\cal D}_{-2-2}+{\cal D}_{22}$ a result that is proportional to
$1+c^2$, which is in agreement with the previously found results for 
the decay of a $\chi_{c2}$ into $J/\psi$ and $\gamma$, equation
(\ref{ForAlp0lim}) in the limit $\alpha=0$.
 Using the expressions for $l_+$ and $l_-$ 
results in
\begin{equation} 
  {\cal D}_{-2-2}+{\cal D}_{22} = 
  |\tilde{c}|^2\left(\frac{M_R^2-M^2}{2}\right)^2M^2
  (1+c^2)\left(1-\frac{1}{2}\left[(s^*c_\phi^*c-c^*s)^2
  +(s^*s_\phi^*)^2\right]\right).
  \label{ForD2222}
\end{equation}

For helicity-2 predictions these results can be compared to the weight
function for the angular distribution, $f_{|\lambda|=2}$, 
presented in \cite{Rhee} and \cite{E760Chic2}
\begin{equation}
  \begin{array}{lcl}
    f_{|\lambda|=2}(\theta,\theta^*,\phi^*) & = &
    \frac{1}{8}A_2^2(1+c^{*2})(1+6c^2+c^4)
    +A_1^2(1-c^{*2})(1-c^4)
  \\
    &&
    +\frac{3}{4}A_0^2(1+c^{*2})(1-2c^2+c^4)
    +\frac{\sqrt{2}}{4}A_2A_1c_\phi^*2s^*c^* s (c^3+3c)
  \\
    &&
    +\frac{\sqrt{6}}{4}A_2A_0(c_\phi^{*2}-s_\phi^{*2}) s^2(1-c^4)
    -\frac{\sqrt{3}}{2}A_1A_0c_\phi^* 2s^*c^* s^3 c.
  \end{array}
  \label{ForFWTasja}
\end{equation}
A pure electric dipole transition would result in amplitudes 
$A_0=0.316$, $A_1=0.548$ and $A_2=0.775$. Experimentally
\cite{E760Chic2} the values
$A_0=0.21$, $A_1=0.49$ and $A_2=0.85$ have been found.
It turns out that the compact expression 
(\ref{ForD2222}) amounts to expression (\ref{ForFWTasja}) with 
the pure electric dipole values for $A_i$.

After the above detailed discussion of a $\chi_{c2}$ decay, we now
briefly mention a similar $\chi_{c1}$ decay, since it may experimentally
contaminate the $\chi_{c2}$ events. Therefore we need a model for the
decay
\begin{equation}
  \chi_{c1}(p_R) \to \gamma(k_1) J/\psi(k_2),
  \label{ForChic1Dec}
\end{equation}
from which one can derive the density matrix. Again, we use the
$\gamma^*\gamma^*$ amplitude where one $\gamma^*$ is replaced by the 
$J/\psi$. We thus have from (\ref{For3P1AmpFor})
\begin{equation}
  {\cal M}(\lambda,\lambda_1,\lambda_2)=\tilde{c}_3M^2\varepsilon[\varepsilon
  (\lambda),\varepsilon^*(\lambda_1),\varepsilon^*(\lambda_2),k_1].
\end{equation}
The density matrix now becomes
\begin{equation}
  {\cal D}_{\lambda\lambda'}=|\tilde{c}_3|^2 M^4 \left\{ \left[ 1 -
  2\frac{(k_1\cdot k_2)}{M^2} \right]
  (k_1 \cdot 
  \varepsilon(\lambda))(k_1 \cdot 
  \varepsilon^*(\lambda'))-\frac{(k_1\cdot k_2)^2}{M^2}(\varepsilon(\lambda)
  \cdot
  \varepsilon^*(\lambda'))\right\}.
\end{equation}
When $k_1$ is again parametrized as (\ref{ForK1ParRRF}), one obtains 
\begin{equation}
  {\cal D}_{\lambda\lambda'}=|\tilde{c}_3|^2M^4
  \left\{\left[1-2\frac{(k_1\cdot k_2)}{M^2}\right]|\vec{k}|^2 u_\lambda
  u_\lambda^* +\frac{(k_1\cdot k_2)^2}{M^2}\delta_{\lambda\lambda'}
  \right\},
  \label{Forchic1decmat}
\end{equation}
where 
\begin{equation}
  u=(u_1,u_0,u_{-1})
  =\left( \frac{s}{\sqrt{2}}e^{i\phi},-c,\frac{s}{\sqrt{2}}e^{-i\phi}
  \right).
\end{equation}

Since in the production the helicity $\pm 1$ states are favoured, the angular
decay distribution will now be approximately of the form 
\begin{equation}
  {\cal D}_{11}+{\cal D}_{-1-1} \sim 1 - \frac{2M^2-M_R^2}{2M^2+M_R^2}c^2 
  \sim 1-0.216 \cdot c^2,
\end{equation}
which has a behaviour opposite to the decay distribution for the
$\chi_{c2}$, e.g. (\ref{ForAlp0lim}). When forward or backward parts of the 
angular distribution $\gamma J/\psi$ are experimentally not accessible
this will affect the $\chi_{c2}$ decay more than the $\chi_{c1}$ decay.
Although the GaGaRes predicted two-photon mediated production of 
$\chi_{c1}\to \gamma J/\psi$ is only 15\% of that
of $\chi_{c2}$ \cite{GaGaResPaper}, 
the $\chi_{c1}$ contribution to the actually
seen $\gamma J/\psi$ events may be higher than this 15\% because of the
opposite decay distribution.
In the experimental determination of the $\gamma^*\gamma^*$ production
through its $\chi_{c2}\to\gamma J/\psi$ decay it would be worthwile 
to estimate the $\chi_{c1}$ contamination of the events.

Such a contamination would somewhat reduce
the really present $\gamma^*\gamma^*\to\chi_{c2}$ production and therefore the 
derived
$\chi_{c2}\to\gamma\gamma$ width. This remark is relevant since at present
the measured 
two-photon width is a factor 3 to 4 higher than the determination from
$p\bar{p}$ collider experiments \cite{PDG}.
More recently it has been argued that this factor is about 2 to 3 
\cite{Patrignani}.


It is clear from these remarks that a precise measurement of the 
$\gamma J/\psi$ angular distribution could clarify the above situation. 
Tools to calculate various density matrices will then be indispensible.

One may wonder whether the above amplitudes predict the ratios of the widths
$\Gamma_1$ and $\Gamma_2$ for the decay processes 
(\ref{ForChic1Dec})
and (\ref{ForResDec}) correctly. This is considered in section 
\ref{SubSecRat}.


\section{Numerical Results}

In this section numerical results for several quantities are given. They
have been obtained by using the event generator GaGaRes \cite{GaGaResPaper},
while extended Galuga results served as a check.

\subsection{Azimuthal distributions}
\begin{figure}[t]
  \begin{center}
    \epsfig{file=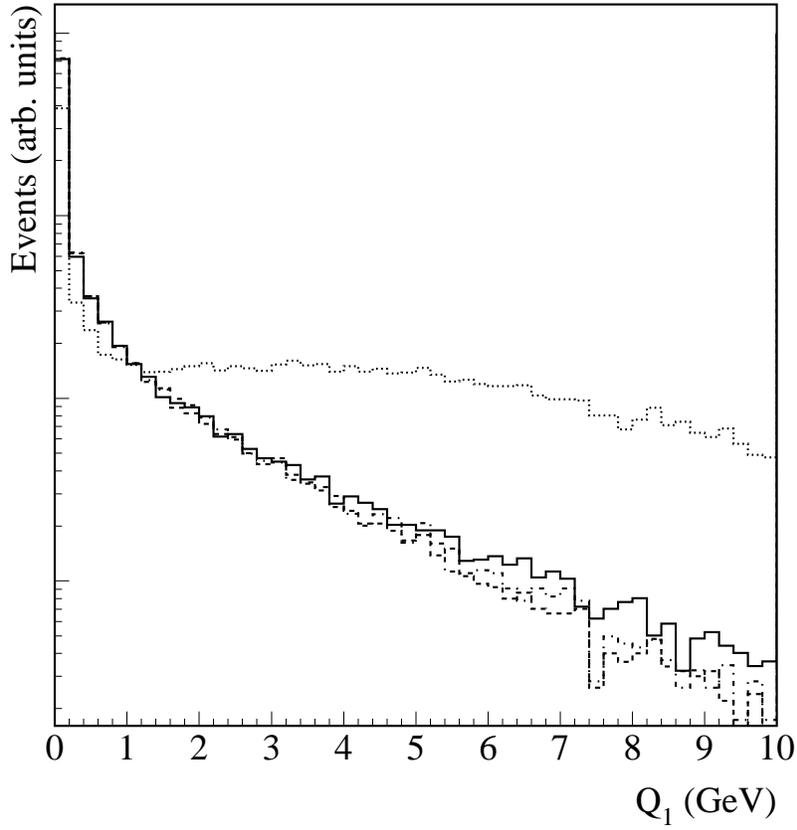, width= 0.8\hsize}
    \caption{{\it The $Q$ spectrum of the photon radiated by the incoming positron for the generated
      $b\bar{b}$ states, the $\eta_b$ (continuous line), the $\chi_{b0}$ 
      (dashed line), the $\chi_{b1}$ (dotted line) and the 
      $\chi_{b2}$ (dash-dotted line).
      The events were
      generated at $\sqrt{s}=$ 190 GeV without cuts on the external particles.
      The vertical scale is logarithmic.
      }}
    \label{FigBBQDep}
  \end{center}
\end{figure}

Before focussing on azimuthal distributions it is useful to show the shape
of the cross section $\mbox{d}\sigma/\mbox{d}Q$ with its marked decrease
with $Q$.
This is displayed in figure \ref{FigBBQDep}, where every curve has an arbitrary
normalization.

\begin{figure}[t]
  \begin{center}
    \begin{tabular}{cc}
      \epsfig{file=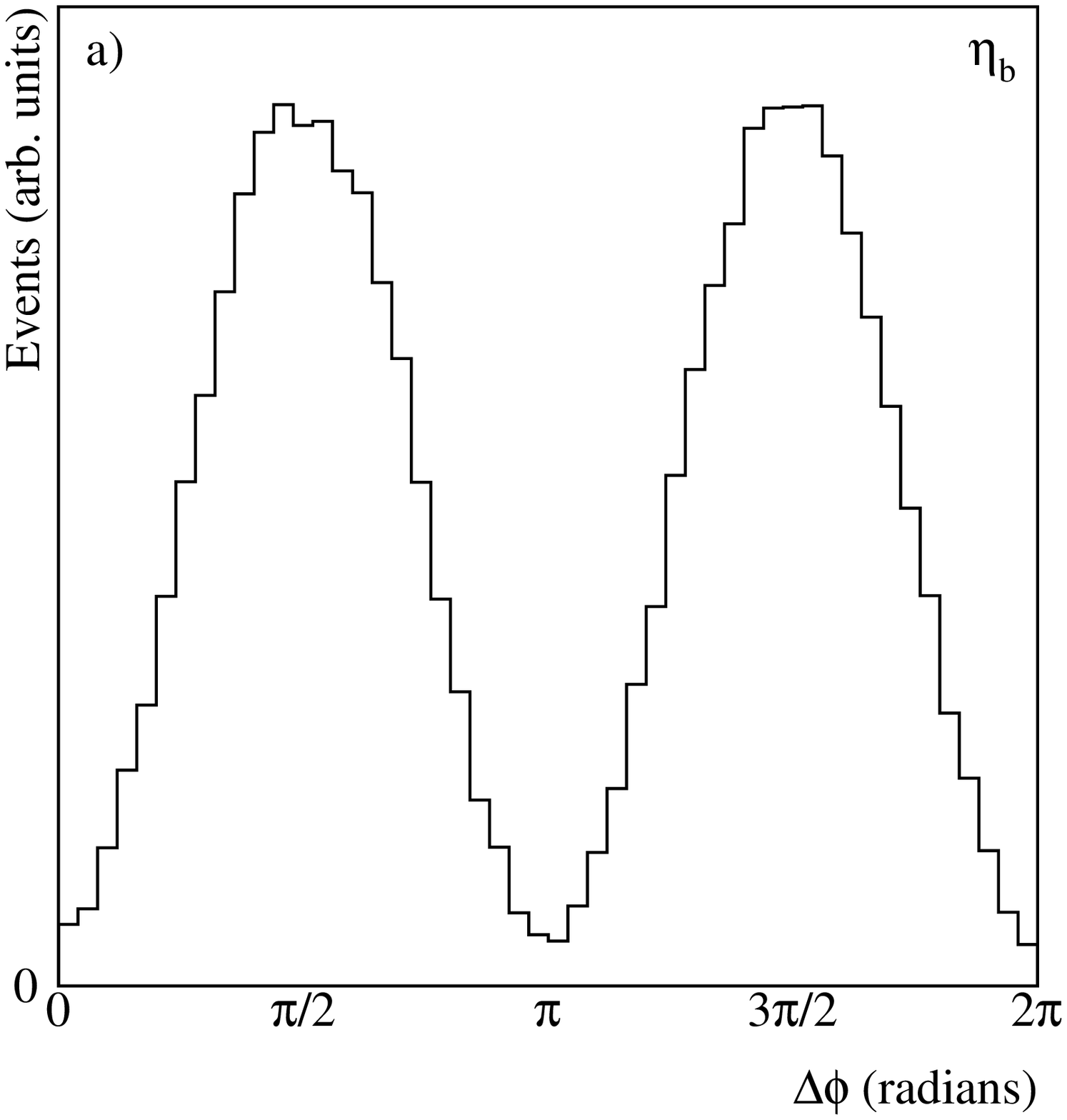, height= 0.5\hsize}
    &
      \epsfig{file=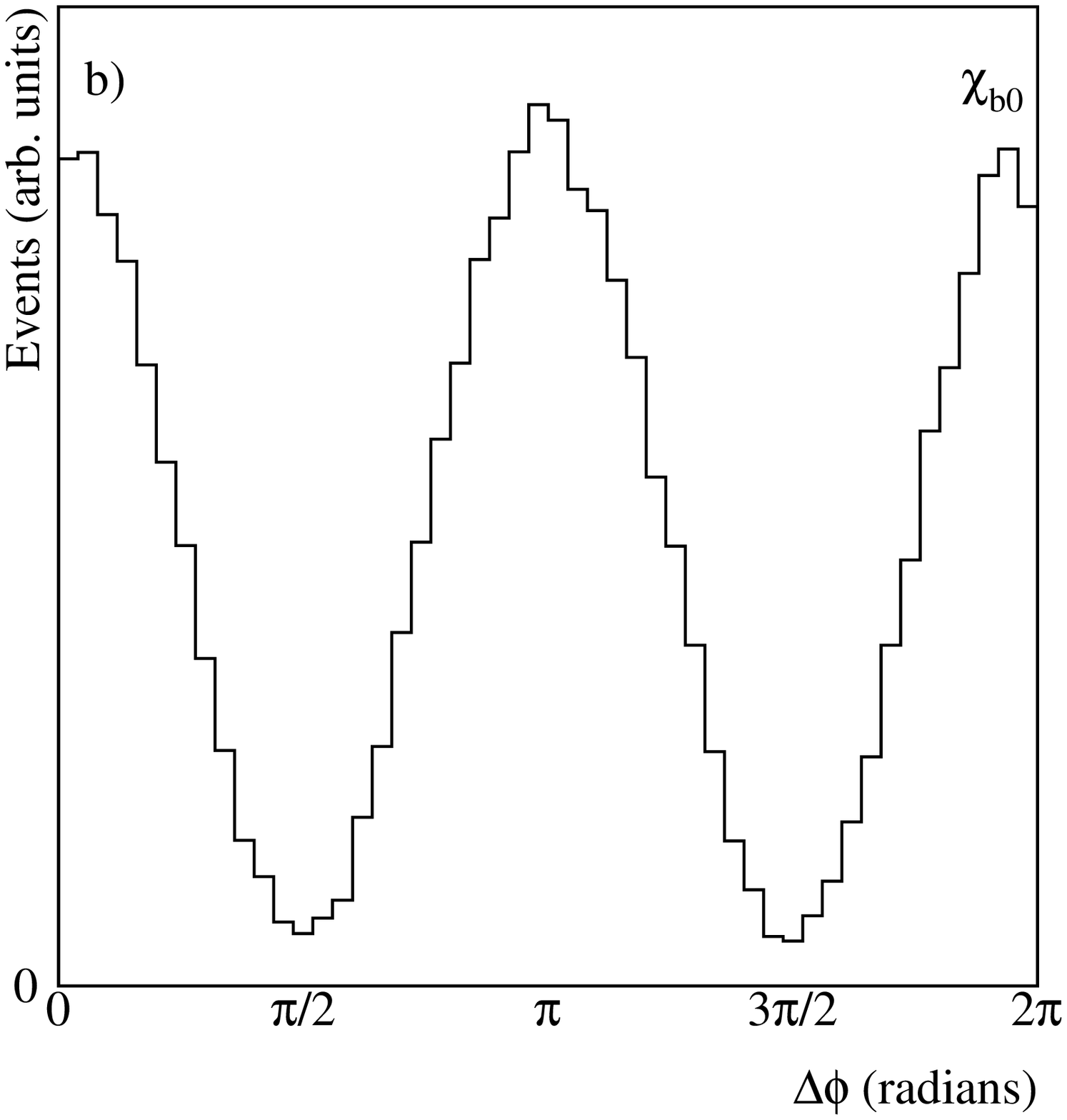, height= 0.5\hsize}
    \\
      \epsfig{file=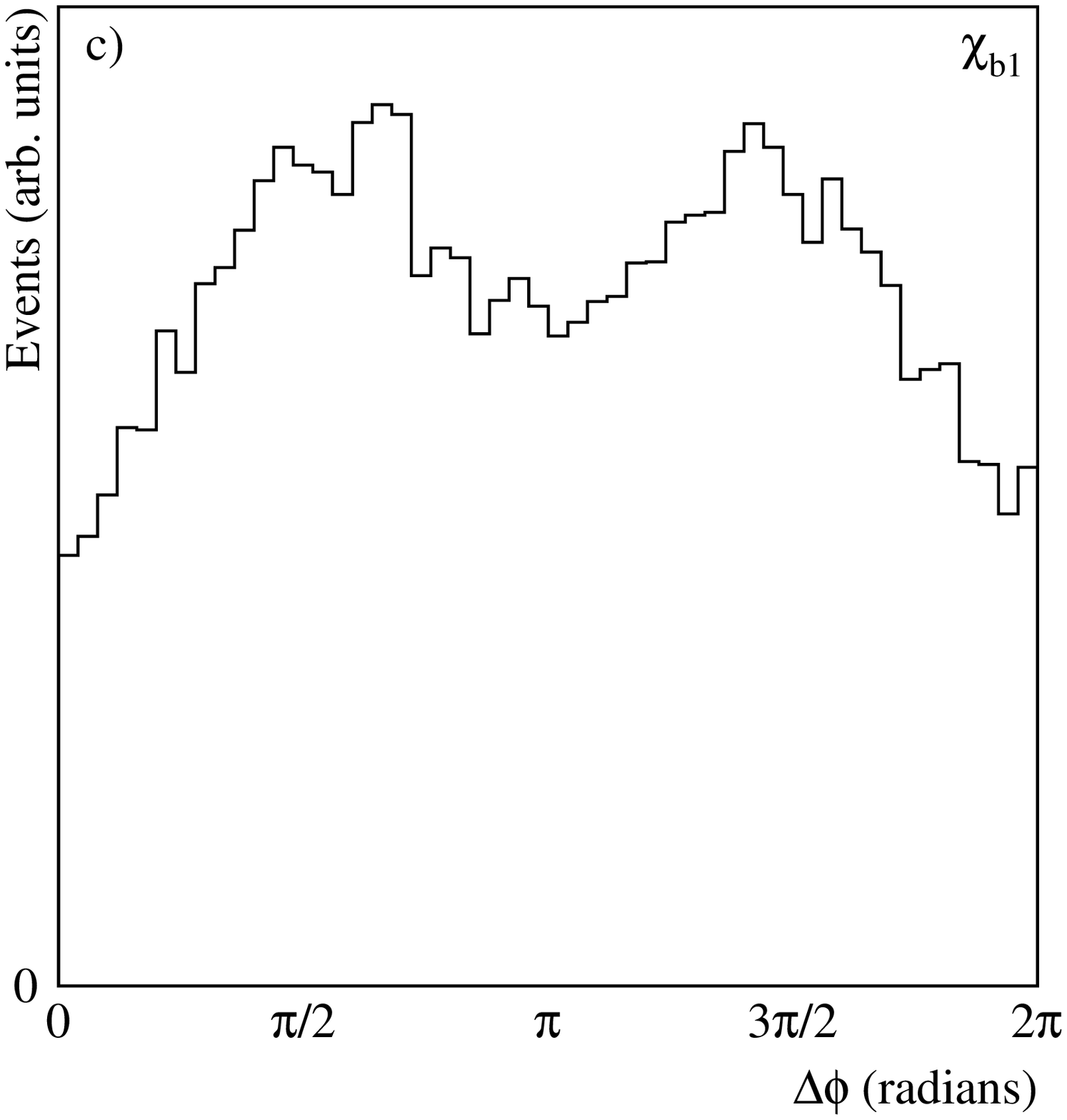, height= 0.5\hsize}
    &
      \epsfig{file=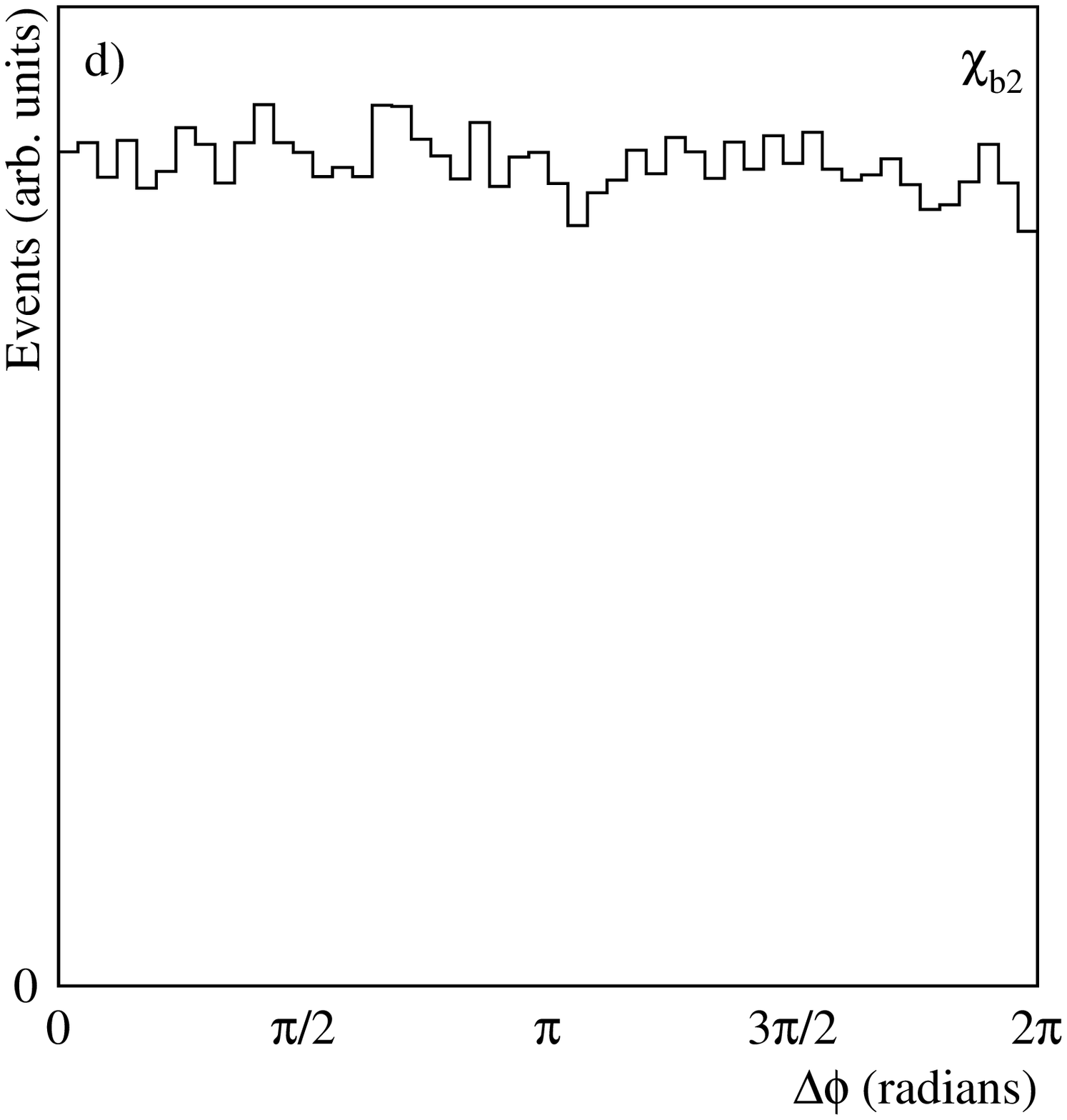, height= 0.5\hsize}
    \end{tabular}
    \caption{{\it The $\Delta \phi$ spectra for the $b\bar{b}$ states
      $\eta_b$ (a), $\chi_{b0}$ (b),
      $\chi_{b1}$ (c) and $\chi_{b2}$ (d). The events have been 
      generated at $\sqrt{s}=190\ \mbox{GeV}$ in the absence of additional
      cuts on the external particles. The vertical scale is linear.}}
      \label{FigDPhiSpecs}
  \end{center}
\end{figure}

Next, results for the $\Delta\phi=\phi_1-\phi_2$ distributions for the
$^1S_0$ and $^3P_J$ bottomonium states are given for 
$\sqrt{s}=190 \ \mbox{GeV}$.
From figures \ref{FigDPhiSpecs}a-\ref{FigDPhiSpecs}d it is seen that 
there are marked differences between distributions for different resonances.
The characteristics can be qualitatively understood when the dominant 
features of the cross sections are considered. 
Similar qualitative arguments, but without numerical results have been
given in the literature \cite{SchulerClose} in connection with
Pomeron production of resonances.
In the first place it
is clear from figure \ref{FigBBQDep} that the largest contributions to
the cross sections come from the region $Q_i^2 \ll M^2$. In those regions
$\Delta \phi \approx \tilde{\phi}$ (cf \cite{BGMS}) such that a 
qualitative understanding of the $\tilde{\phi}$ distribution in the 
BGMS cross section formula (\ref{ForBGMS}) is sufficient to explain
the features of figures \ref{FigDPhiSpecs}a-\ref{FigDPhiSpecs}d. The 
behaviour of the BGMS formula for the various resonances now follows
from the form factors $f_{AB}$ and $g_{AB}$ in table \ref{TabFFs1}
on page \pageref{TabFFs1} and the small $Q_i^2$ approximation.
One then obtains $\tilde{\phi}$ distributions, where different 
photon density matrices play a role. Using the approximate numerical 
relations valid at $\sqrt{s}=190 \ \mbox{GeV}$ and in the
small $Q_i^2$ region
\begin{equation}
  \rho^{++} \approx \rho^{+-}, \hspace{1.0cm}
  \rho^{+0} \approx 1.5 \rho^{++}, \hspace{1.0cm}
  \rho^{00} \approx 2.3 \rho^{++},
\end{equation}
one arrives at the following shapes for the $\tilde{\phi}$ distributions:

\noindent
$ \eta_b:$
\begin{equation}
 \begin{array}{lcl} 
           \frac{{d}\sigma}{d\tilde{\phi}} & \sim &
           \kappa\frac{8 X}{\nu^2} \left(
           \rho_1^{++} \rho_2^{++} - \rho_1^{+-} \rho_2^{+-} \cos 
           (2\tilde{\phi}) \right)
  \\      & 
           \sim & \kappa\frac{16 X}{\nu^2} \rho_1^{++} \rho_2^{++}
           \sin^2 \tilde{\phi},
  \end{array}
\end{equation}
$ \chi_{b0}:$
\begin{equation}
 \begin{array}{lcl} 
              \frac{{d}\sigma}{d\tilde{\phi}} & \sim &
              4\kappa\left(\frac{X+\nu M^2}{3\nu^2}\right)\left(
              \rho_1^{++} \rho_2^{++} + \rho_1^{+-} \rho_2^{+-} \cos 
              (2\tilde{\phi}) \right)
  \\        &
	      \sim &
              8\kappa\left(\frac{X+\nu M^2}{3\nu^2}\right)
              \rho_1^{++} \rho_2^{++}\cos^2\tilde{\phi},
  \end{array}
\end{equation}
$ \chi_{b1}:$
\begin{equation}
 \begin{array}{lcl} 
              \frac{{d}\sigma}{d\tilde{\phi}} & \sim &
              4\kappa\left( \frac{Q_1^2-Q_2^2}{2\nu}\right)^2
              \left( \rho_1^{++} \rho_2^{++} - \rho_1^{+-} \rho_2^{+-} \cos 
              (2\tilde{\phi}) \right) \\
            &&
              +
              4\kappa\left( \frac{M}{2\nu}\right)^2\left(
              \rho_1^{++}\rho_2^{00} Q_2^2 + \rho_1^{00} \rho_2^{++} Q_1^2
              -2 \rho_1^{+0} \rho_2^{+0} Q_1Q_2 \cos \tilde{\phi}\right)
  \\        &
              \sim &
              \frac{2\kappa}{\nu^2} \rho_1^{++} \rho_2^{++}
              \left( M^2(Q_1^2+Q_2^2)-2M^2Q_1Q_2\cos\tilde{\phi}
              +(Q_1^2-Q_2^2)^2\sin^2\tilde{\phi}\right),
  \end{array}
\end{equation}
$\chi_{b2}:$
\begin{equation}
 \begin{array}{lcl} 
   \frac{{d}\sigma}{d\tilde{\phi}} & \sim &
   \kappa \left(\frac{M^2}{2\nu}\right)^2.
 \end{array}
\end{equation}
It is clear that the $\sin^2\tilde{\phi}$ and $\cos^2\tilde{\phi}$ and constant
distributions show up in figures \ref{FigDPhiSpecs} a,b and d. A more
complex structure arises in figure \ref{FigDPhiSpecs} c, but also here the
$\cos \tilde{\phi}$ and $\sin^2 \tilde{\phi}$ distributions can be 
recognized on a constant background.

The $^1D_2$ $b\bar{b}$ state has not
been included in the plot as it would lie exactly on top of the plot for the
$^1S_0$ $b\bar{b}$ state.

For completeness we note that for the $c\bar{c}$ states a similar 
$\Delta\phi$ behaviour has been found.

\subsection{Density matrices for resonance production}

\begin{figure}[tb]
  \begin{center}
  \begin{tabular}{cc}
    \epsfig{file=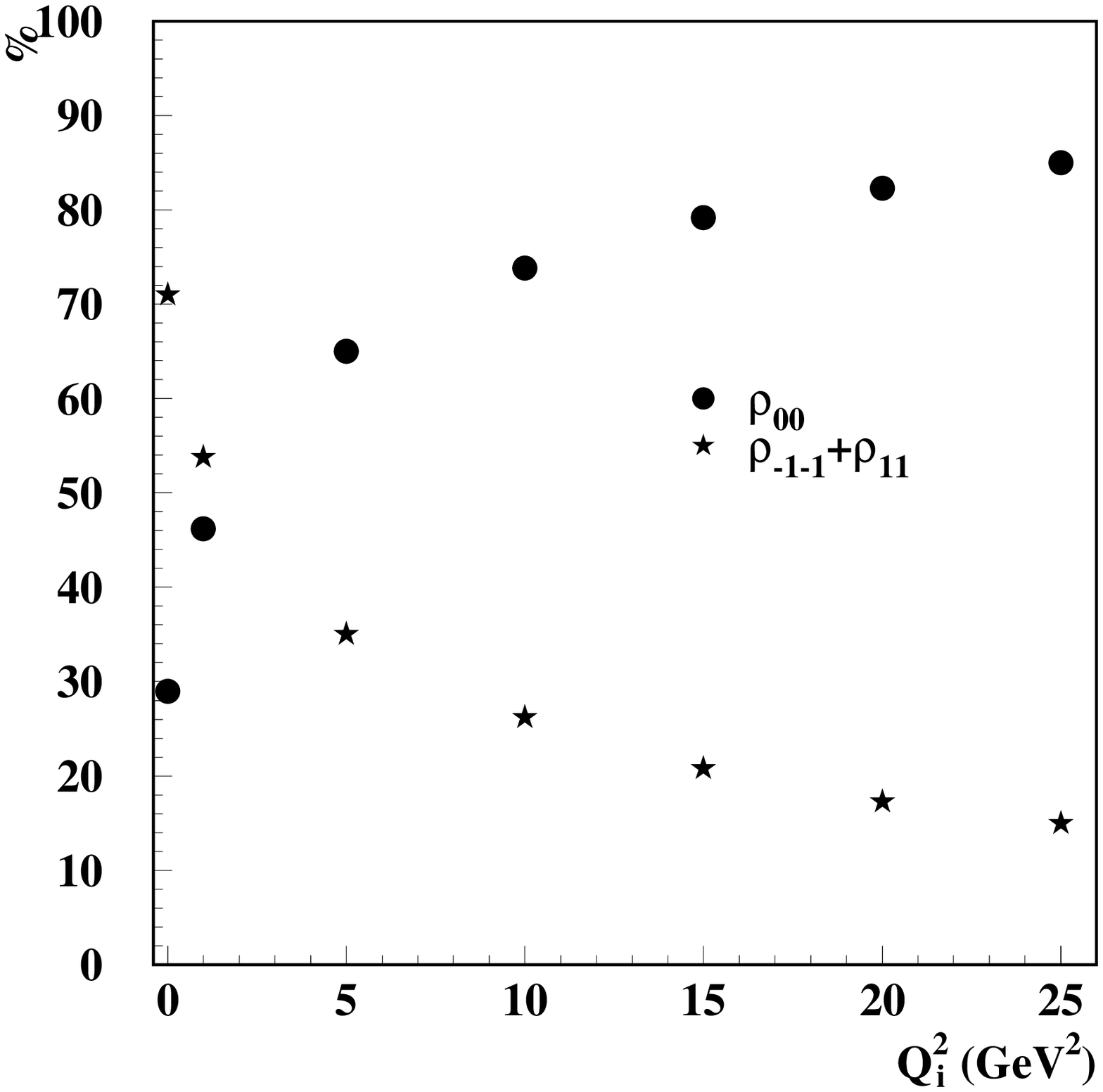, width= 0.45\hsize}
  &
    \epsfig{file=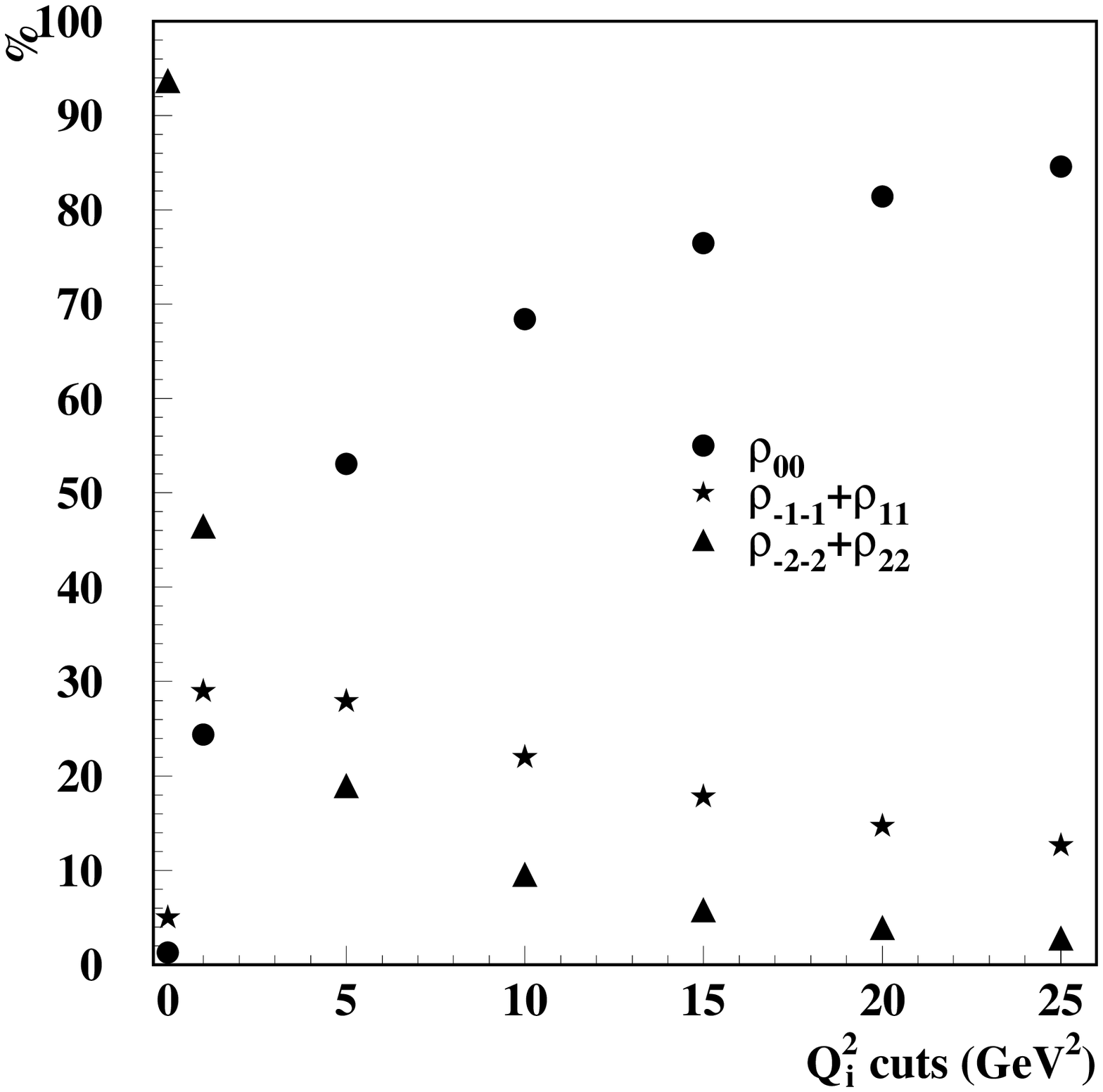, width= 0.45\hsize}
  \end{tabular}
  \caption{{\it Contributions of the different diagonal density matrix 
  elements of the $\chi_{c1}$ resonance
  (left plot) and the $\chi_{c2}$ resonance (right plot) to the total 
  cross section at $\sqrt{s}=91.5 \ \mbox{GeV}$, for different $Q_i^2$ cuts.
  (I.e. $Q_i^2 > 0,1,5,\ldots,25 \ \mbox{GeV}^2$.)
  }}
  \label{Figchic1chic2helcon}
  \end{center}
\end{figure}

For the $c\bar{c}$ resonances $\chi_{c2}$ and $\chi_{c1}$ 
we have studied how the different helicities of the resonance contribute
to the cross section as a function of the $Q_i^2$ cuts. 
The results for the diagonal elements in the 
RRF
are given in 
figure \ref{Figchic1chic2helcon}
In these plots both $Q_1^2$ and $Q_2^2$ 
have to be greater than the value given on 
the horizontal axis. 
From the right plot in 
figure \ref{Figchic1chic2helcon} one can see that in the absence of cuts 
the dominant contribution comes from the helicity-2 components. This is in
agreement with \cite{L3Braccini}
where a similar 
dominance was found for the helicity-2 component for the $f_2'(1525)$ 
resonance.
For high $Q_i^2$ cuts the helicity-0 component starts to dominate.
In the BGMS-formalism \cite{BGMS} 
this contribution comes from the $\sigma_{SS}$
component that is proportional to $Q_1^2Q_2^2$.
This result also agrees with theoretical predictions found by
Close \cite{Close}, in which he states that 
the helicity-0 contribution is of the order ${\cal O}(tt'M^2)$.
From the left plot in figure \ref{Figchic1chic2helcon}
one can see that in the case of the 
$\chi_{c1}$ resonance in the absence of cuts the helicity-1 component
is dominant. It would be interesting to verify this statement with 
experimental data. For higher $Q_i^2$ cuts again the helicity-0 component 
 starts to dominate.

\subsection{Density matrix for the decay $\chi_{c2}\to \gamma J/\psi$}
\begin{figure}[tb]
  \begin{center}
  \epsfig{file=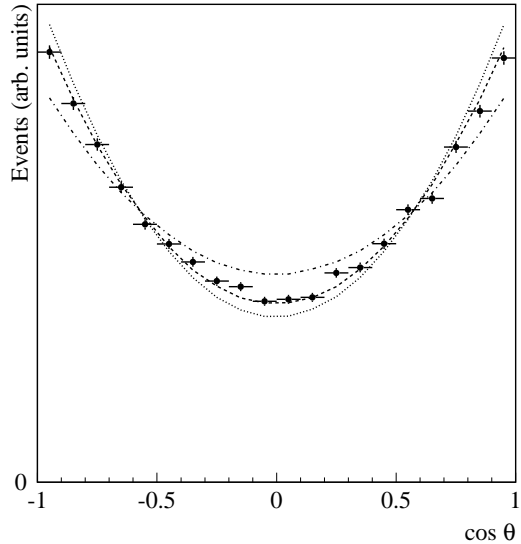, width= 0.53\hsize}
  \end{center}
  \caption{{\it The distribution of the polar angle of the outgoing photon
    in the RRF. The data points show the result from the calculation using
    the combined matrix element both for the production and decay. The dashed
    line shows the distribution (\ref{ForForXXX})
    for the complete density matrix for the
    decay of the resonance where the production is purely helicity-2. The
    dotted line represents the distribution (\ref{ForForYYY}) 
    according to the pure 
    dipole transition and the dashed-dotted line shows the distribution
    that has been experimentally observed \protect\cite{E760Chic2}. The
    vertical scale is linear.
    }}
 \label{FigCosThDisTotMats}
\end{figure}

In section 5 we have constructed the density matrix for the 
complete decay of a $\chi_{c2}$ resonance into a $J/\psi$ and a photon,
where the $J/\psi$ subsequently decays into electrons or muons.
We can now use equation 
(\ref{ForTraceDensMatDef})
to combine the density matrix for the 
two-photon production 
of a $\chi_{c2}$ resonance
with the density matrix 
for the complete decay of the resonance to get the full matrix element 
squared. 
The average production density matrix is given in table 3. Again it is clear
that the helicity-2 diagonal elements are dominant.
The combined production and decay 
matrix elements (\ref{ForDecMatTot}) 
have been used to generate the distribution
of the angle between the outgoing photon and the boost direction in the
RRF. The results are given as data points
in figure \ref{FigCosThDisTotMats}. 

In this figure
also 
results 
are included which are obtained from the assumption of pure helicity-2
production.
Under this assumption the ${\cal D}_{22}+{\cal D}_{-2-2}$ distributions 
(\ref{ForForXXX}), (\ref{ForForYYY})
or (\ref{ForExpDis22}) can be chosen. Comparing the data points of the 
full calculation and distribution 
(\ref{ForForXXX})
shows that the
production is indeed helicity-2 dominated. Comparing the distributions
(\ref{ForForYYY})
and 
(\ref{ForExpDis22}) with the model distribution
(\ref{ForForXXX})
shows that the model is closer to experiment than the
simple dipole distribution.

\renewcommand{\arraystretch}{1.2}
\begin{table}[t]
  \[
    \begin{array}{|c|c|c|c|c|c|} \hline
       & \lambda'=+2 &  \lambda'=+1 &  \lambda'=0 &  \lambda'=-1 & \lambda'=-2 
    \\ \hline
    \lambda=+2 & 0.47 & 2.0\cdot 10^{-4} & 0.012 & 6.0 \cdot 10^{-4} & 0.029
    \\ 
          &           & 4.72           & 6.27  & 5.56              & 0.13
    \\ \hline
    \lambda=+1  &      & 0.022          & 5.4 \cdot 10^{-4} & 4.1 \cdot 10^{-3}
          &   6.0 \cdot 10^{-4} \\
          &           &                & 1.53              & 0.023 
          &   2.41 
    \\ \hline
    \lambda=0   &      &      &  0.012 &  5.4 \cdot 10^{-4} &  0.012 
    \\ 
                &      &      &        &  4.67              & 6.27
    \\ \hline
    \lambda=-1  &      &      &        &  0.022             &  2.0\cdot 10^{-4}
    \\
                &      &      &        &                    & 1.58
    \\ \hline
    \lambda=-2  &      &      &        &                    & 0.47
    \\
                &      &      &        &                    &
    \\ \hline
    \end{array}
  \]
  \caption{{\it The normalized density matrix $\rho_{\lambda\lambda'}=
  |\rho_{\lambda\lambda'}|e^{i\phi_{\lambda\lambda'}}$
    for $\chi_{c2}$ production at
    $\sqrt{s}=190 \ \mbox{GeV}$ in the absence of cuts on the external 
    particles. In each entry the absolute value $ |\rho_{\lambda\lambda'}|$
    of the density matrix
    (upper value) and the phase $\phi_{\lambda\lambda'}$ 
    (lower value) are given. The open boxes are filled by Hermitian 
    conjugation of the other elements.
    }}
  \label{TabDensMat}
\end{table}

\subsection{The ratio $\Gamma(\chi_{c1}\to \gamma J/\psi)/
\Gamma(\chi_{c2}\to \gamma J/\psi)$}
\label{SubSecRat}
As the trace of the density matrices for the decay of a $\chi_{cJ}$ 
resonance into a photon and a $J/\psi$ yields the matrix element squared
for this process, we can use equations (\ref{ForDecMatTot}) and 
(\ref{Forchic1decmat}) to calculate the ratio of the two decay widths.
We assume $\tilde{c}_3/\tilde{c}_4=c_3/c_4$, i.e. we assume that the
replacement of a $\gamma^*$ by a $J/\psi$ introduces in both matrix elements
(\ref{For3P1AmpFor}) and (\ref{For3P2Exp}) the same scale factor. 
The mass of the 
$\chi_{cJ}$ particle is denoted by $M_{R_J}$.
Using the expressions for the traces of (\ref{ForDecMatTot}) and
(\ref{Forchic1decmat}) yields for the ratio
\begin{equation}
  R_{th}=\frac{\Gamma(\chi_{c1}\to \gamma J/\psi)}
  {\Gamma(\chi_{c2}\to \gamma J/\psi)}
  =5\frac{M^2(M_{R_1}^2+M^2)}{M_{R_1}^4}\left(
  \frac{M_{R_2}}{M_{R_1}}\right)^4 \left(\frac{\alpha}{\beta}\right)
  \frac{1}{10-5\alpha+\alpha^2},
\end{equation}
where $\beta$ has been defined 
like $\alpha$
\begin{equation}
  \beta=1-\frac{M^2}{M_{R_1}^2}\approx 0.225.
\end{equation}
Inserting the numerical values gives
\begin{equation}
R_{th} \approx 0.8969.
\end{equation}


This number should be compared to the experimentally measured 
\cite{PDG} ratio
\begin{equation}
  R_{exp}
  = 
  \frac{\Gamma_{tot}(\chi_{c1})\mbox{Br}(\chi_{c1}\to\gamma J/\psi)}
  {\Gamma_{tot}(\chi_{c2})\mbox{Br}(\chi_{c2}\to\gamma J/\psi)}
  =
  0.89 \pm 0.15,
\end{equation}
where in the calculation of the error the different contributions to this
error are taken
to be independent. Using the data selection of \cite{Patrignani}
this number becomes
\begin{equation}
  R_{exp}=0.76\pm0.26.
\end{equation}
These experimental numbers are compatible with the above theoretical 
prediction.

\section*{Acknowledgements}
The authors would like to thank Dr. G. Schuler for making the program 
Galuga available and providing detailed information on it.


\appendix

\section{Transforming the density matrices}
In section 4
we have constructed the complete 
density matrix in the
BGMS frame. 
However from an experimental point of view the RRF is a more convenient
reference frame.
As the elements of the density matrix depend on the chosen polarization
vectors/tensors, the density matrix is frame dependent. When one wants
to rotate the quantization axis over the angles $\theta_p$ and $\phi_p$
the density matrix $\rho$ changes into
the density matrix $\tilde{\rho}$ according to
\begin{equation}
  \tilde{\rho}_{\lambda\lambda'}=
  A_{\lambda\mu}A_{\lambda'\mu'}^*\rho_{\mu\mu'}.
  \label{ForTransDef}
\end{equation}
Thus a density matrix $\rho$ calculated in the BGMS frame can be 
transformed to the RRF. In the BGMS frame the boost direction which
was needed to get to the $\chi_{c2}$ rest system is 
characterized by $\theta_p$ and $\phi_p$.

For spin-1 resonances this transformation matrix $A$ can be shown to be
\renewcommand{\arraystretch}{1.5}
\begin{equation}
  A_{\lambda\mu}=
  \left(
    \begin{array}{ccc}
      \frac{1+c}{2}\fm   & \frac{s}{\sqrt{2}}   & \frac{1-c}{2}\fp   \\
       -\frac{s}{\sqrt{2}}\fm  & c              & \frac{s}{\sqrt{2}}\fp    \\
       \frac{1-c}{2}\fm  & -\frac{s}{\sqrt{2}}  &\frac{1+c}{2}\fp                 
    \end{array}
  \right)
  \label{transa}
\end{equation}
The order of the indices $\lambda,\mu$ is taken to be ($+$, $0$, $-$).
For spin-2 resonances this transformation matrix reads (with
order $+2$, $+1$, $0$, $-1$, $-2$)
\begin{equation}
  A_{\lambda\mu}=
  \left(
  \begin{array}{ccccc}
    \left(\frac{1+c}{2}\right)^2 \emtp &
    \frac{(1+c)s}{2}\emp &
    \frac{s^2}{2}\sqrt{\frac{3}{2}} &
    \frac{(1-c)s}{2}\ep &
    \left(\frac{1-c}{2}\right)^2 \etp 
  \\
    -\frac{(1+c)s}{2}\emtp &
    \frac{(1+c)(2c-1)}{2}\emp &
    cs\sqrt{\frac{3}{2}} &
    \frac{(1-c)(2c+1)}{2}\ep &
    \frac{(1-c)s}{2}\etp 
  \\
    \frac{1}{2}\sqrt{\frac{3}{2}}s^2\emtp &
    -\sqrt{\frac{3}{2}}sc\emp &
    \frac{3c^2-1}{2} &
    \sqrt{\frac{3}{2}}sc\ep &
    \frac{1}{2}\sqrt{\frac{3}{2}}s^2\etp 
  \\
    -\frac{(1-c)s}{2}\emtp &
    \frac{(1-c)(2c+1)}{2}\emp &
    -cs\sqrt{\frac{3}{2}} &
    \frac{(1+c)(2c-1)}{2}\ep &
    \frac{(1+c)s}{2}\etp 
  \\
    \left(\frac{1-c}{2}\right)^2 \emtp &
    -\frac{(1-c)s}{2}\emp &
    \frac{s^2}{2}\sqrt{\frac{3}{2}} &
    -\frac{(1+c)s}{2}\ep &
    \left(\frac{1+c}{2}\right)^2 \etp
  \end{array}
  \right)
  \label{transmata}
\end{equation}
In the expressions for the transformation matrices we have introduced
the shorthand notation
\begin{equation}  
  c=\cos (\theta_p), s=\sin (\theta_p), \phi=\phi_p.
  \label{ForShortHand}
\end{equation}



\clearpage

\end{document}